\documentclass[showpacs,amsmath,amssymb,prd,nofootinbib,twocolumn,floatfix]{revtex4}
\usepackage{graphicx}% Include figure files
\usepackage{dcolumn}% Align table columns on decimal point
\usepackage{bm}% bold math
\usepackage{epsf}
\usepackage{hyperref}
\usepackage{amsfonts}
\usepackage{verbatim}
 
\begin{document}
 
\title{Trumpet slices of the Schwarzschild-Tangherlini spacetime}

\author{Kenneth A. Dennison}

\affiliation{Department of Physics and Astronomy, Bowdoin College,
Brunswick, Maine 04011}

\author{John P. Wendell}

\affiliation{Department of Physics and Astronomy, Bowdoin College,
Brunswick, Maine 04011}

\author{Thomas W. Baumgarte}

\altaffiliation{Also at Department of Physics, University of Illinois at
        Urbana-Champaign, Urbana, IL, 61801}

\affiliation{Department of Physics and Astronomy, Bowdoin College,
Brunswick, Maine 04011}

\author{J. David Brown}

\affiliation{Department of Physics, North Carolina State University, Raleigh, North Carolina 27695}

\date{{\rm draft of} \today}

\begin{abstract}
We study families of time-independent maximal and $1+\rm{log}$ foliations of the Schwarzschild-Tangherlini spacetime, the spherically-symmetric vacuum black hole solution in $D$ spacetime dimensions, for $D\geq 4$.  We identify special members of these families for which the spatial slices display a trumpet geometry.   Using a generalization of the $1+\rm{log}$ slicing condition that is parametrized by a constant $n$ we recover the results of Nakao, Abe, Yoshino and Shibata in the limit of maximal slicing. We also construct a numerical code that evolves the BSSN equations for $D=5$ in spherical symmetry using moving-puncture coordinates, and demonstrate that these simulations settle down to the trumpet solutions.  
%Finally, we speculate on why these trumpet solutions serve as attractors in dynamical solutions even in maximal slicing.
\end{abstract}


\maketitle

\section{Introduction}
Numerical relativity simulations of binary black holes have matured dramatically in recent years.  
Starting with the first complete simulations of binary black hole mergers \cite{Pre05b,CamLMZ06,BakCCKM06a}, a large number of papers on mergers of binaries with varying mass ratios and black hole spins have appeared.  Some of the results of these simulations, including the surprisingly large recoil speed of the merger remnant for certain spin orientations (e.g.~\cite{CamLZM07b,GonHSBH07}), have also triggered numerous studies of the astrophysical consequences of these findings.

Many numerical simulations of black holes adopt the Baumgarte-Shapiro-Shibata-Nakamura (BSSN) formulation of Einstein's equations \cite{ShiN95,BauS99} together with moving-puncture coordinates \cite{CamLMZ06,BakCCKM06a} (see also \cite{BauS10}).  The latter consist of the 1+log slicing condition for the lapse \cite{BonMSS95} and the $\bar \Gamma$-driver gauge condition for the shift \cite{AlcBDKPST03}.  The role of these coordinates in stabilizing the numerical simulations has been clarified by considering the late-time behavior of spatial slices of the Schwarzschild spacetime when evolved with moving-puncture coordinates \cite{HanHPBO06,Bro08,HanHOBO08}.  In particular, these studies showed that, at late times, these slices form a ``trumpet" geometry, meaning that the slices asymptotically approach a finite areal radius and never reach the spacetime singularity (see Fig.~2 in \cite{HanHOBO08} for an embedding diagram that motivates the name of this geometry).

Several groups have also started to simulate black holes in higher dimensions (e.g.~\cite{ChoLOPPV03, GarLP05, LehP10, YosS09, NakAYS09, ShiY10a, ShiY10b, ZilWSCGHN10, WitZGCHNS10}).  Some of these studies aim at exploring the rich geometric structure of black holes in higher dimensions (see, e.g., \cite{EmpR08}), while others are motived by speculations that miniature black holes might be created in high-energy collisions in particle colliders (e.g.~the LHC), a scenario that requires additional spacelike dimensions (see, e.g.,~\cite{Kan09} for a review).

Motivated by the success of the BSSN formalism with moving-puncture coordinates in $D=4$ spacetime dimensions, several of the above simulations adopt this method for $D > 4$ as well.  In this paper we analyze the late-time behavior of spatial slices of spherically symmetric solutions in $D\geq 4$ spacetime dimensions when they are evolved with moving-puncture coordinates.  More specifically, we construct stationary $1+\log$ slices of the Schwarzschild-Tangherlini spacetime \cite{Tan63}, the spherically symmetric vacuum black hole solution for $D \geq 4$.   We demonstrate that, as for $D=4$, these slices display a trumpet geometry.  Parameterizing the $1+\log$ slicing condition with a parameter $n$, we show that we recover the results of \cite{NakAYS09} for maximal slicing in the limit $n\to\infty$.  We also recover the results of \cite{HanHOBO08} for $D = 4$.  Finally, we perform numerical simulations for $D=5$ and demonstrate that these simulations indeed settle down to the trumpet slices.

This paper is organized as follows.   In Section \ref{SchwTanSection}, we introduce the Schwarzschild-Tangherlini spacetime \cite{Tan63} and introduce a ``height function" that we will use for coordinate transformations in the following sections.  In Section \ref{MaxSlicSection}, we construct a family of maximal slices of the Schwarzschild-Tangherlini spacetime, reproducing earlier results of \cite{NakAYS09}, and identify a special member of this family that displays a trumpet geometry.  In Section \ref{1pluslogSection}, we derive the main result of this paper, a stationary $1+\rm{log}$ trumpet solution for the Schwarzschild-Tangherlini spacetime.  We recover the corresponding $D=4$ results from \cite{HanHOBO08} and the maximal slicing results of \cite{NakAYS09} in the appropriate limits.  In Section \ref{NumWorkSection} we perform dynamical simulations for $D=5$ in spherical symmetry and demonstrate that, at late times, these simulations indeed settle down to the trumpet slices derived in the previous sections.  We conclude with a brief summary in Section \ref{Summary}.

Throughout this paper we adopt units in which the speed of light is unity, $c=1$.  However, we do not set to unity the gravitational constant $G$ of a $D$-dimensional spacetime, since keeping $G$ in the analytical expressions makes their units more transparent.

%=====================================================================

\section{Schwarzschild-Tangherlini}
\label{SchwTanSection}

The generalization of the Schwarzschild spacetime, which describes spherically symmetric vacuum solutions in $D=4$ spacetime dimensions, to higher dimensions $D$ is the Schwarzschild-Tangherlini solution \cite{Tan63}.  Adopting the notation of \cite{EmpR08}, this solution can be written as
\begin{equation}
\label{STle}
ds^{2} = - f_0 dt^2+ f_0^{-1} dR^2 +R^{2}d\Omega^{2}_{D-2},
\end{equation}
where we have defined
\begin{equation}
\label{deff0}
f_{0}\left(R\right) = 1 - \frac{\mu}{R^{D-3}},
\end{equation}
and where the mass parameter $\mu$ is given by
\begin{equation}
\label{massparameter}
\mu = \frac{16\pi G M}{\left(D-2\right)\Omega_{D-2}}.
\end{equation}
The area of a unit $D-2$ sphere is
\begin{equation}
\label{areaunitdminus2sphere}
\Omega_{D-2} = \frac{2 \pi^{(D-1)/2}}{\Gamma\left( (D-1)/2 \right)},
\end{equation}
and the line element on this sphere is 
\begin{equation}
\label{defdOmegasqu}
d\Omega^{2}_{D-2}=d\chi_{2}^{2}+\sin^{2}\chi_{2}d\chi_{3}^{2}+\dots+
	\left(\prod_{l=2}^{D-2}\sin^{2}\chi_{l} \right) d\chi_{D-1}^{2},
\end{equation}
where the $\chi_l$ (with $l = 2, 3,\ldots, D-1$) are the angles on the $D-2$ sphere.  The angles $\chi_2$ and $\chi_{D-1}$ are often called $\theta$ and $\phi$.  We also point out that $GM$ has dimensions of $(length)^{D-3}$.  For $D=4$ we recover $\mu = 2 GM$ and $\Omega_2 = 4 \pi$, as expected.

In the line element (\ref{STle}), $R$ is the generalization of the Schwarzschild (or areal) radius.  In numerical applications, it is often convenient to express the metric in terms of an isotropic radius $r$, so that the spatial part of the metric can be transformed to cartesian coordinates very easily.  To do so, we 
set the spatial part of the line element (\ref{STle}) equal to its isotropic counterpart,
\begin{equation}
\label{toisotropic}
\frac{dR^{2}}{1 - \mu/R^{D-3}} + R^{2}d\Omega^{2}_{D-2} = \psi^{4/(D-3)}\left(dr^{2} + r^{2}d\Omega^{2}_{D-2}\right),
\end{equation}
where the exponent on the conformal factor $\psi$ has been chosen for convenience.  From the identification (\ref{toisotropic}) we see that
\begin{equation}
\label{relateRr}
R^{2} = \psi^{4/(D-3)} r^{2},
\end{equation}
and
\begin{equation}
\label{relatedRdr}
\frac{dR^{2}}{1 - \mu/R^{D-3}} = \psi^{4/(D-3)}dr^{2}.
\end{equation}
Eliminating $\psi$ we now find
\begin{equation}
\label{defrintegral}
\pm\int\frac{dr}{r} % =  \int\frac{1}{\sqrt{1 - \mu/R^{D-3}}}\frac{dR}{R} 
= \int\frac{R^{D-3}dR}{\sqrt{R^{2\left(D-2\right)}-\mu R^{D-1}}}.
\end{equation}
Integrating both sides of the equation we obtain \cite{KolSP04}
\begin{equation}
r=\mu^{1/(D-3)}\left(\frac{\sqrt{R^{D-3}/\mu - 1}+\sqrt{ R^{D-3}/\mu}}{2}\right)^{2/(D-3)}
\end{equation}
or, solving for $R$,
\begin{equation}
R=r\left(1+\frac{\mu}{4r^{D-3}}\right)^{2/(D-3)}.
\end{equation}
Inserting this into (\ref{relateRr}) we find the conformal factor
\begin{equation} \label{ST_conf_factor}
\psi=1+\frac{\mu}{4r^{D-3}}.
\end{equation}
As expected, these results reduce to the usual Schwarzschild expressions for $D=4$.  In Section \ref{NumWorkSection} we will adopt the above expressions for $D=5$ as initial data.

In order to explore alternative slicings (or foliations) of the Schwarzschild-Tangherlini spacetime in Sections \ref{MaxSlicSection} and \ref{1pluslogSection} we now introduce a new time coordinate 
\begin{equation}
\label{deftbarheightfunc}
\bar{t} = t + h\left(R\right),
\end{equation}
where $h\left(R\right)$ is the height function \cite{Rei73,BauS10}.  As we will see in the following Sections, different slicing conditions result in different ordinary differential equations for $h\left(R\right)$.  With the new time coordinate, the line element (\ref{STle}) becomes
\begin{equation}
\label{STlecompactnotnew}
ds^{2} = -f_{0}d\bar{t}^{2} + 2f_{0}h'd\bar{t}dR + \frac{1-f_{0}^{2}h'^{2}}{f_{0}}dR^{2} + R^{2}d\Omega^{2}_{D-2}
\end{equation}
where $h'(R) \equiv dh/dR$.

We can write a general line element in $\left(D-1\right)+1$ form as
\begin{equation}
\label{dminus1plus1le}
ds^{2} = -\alpha^{2}dt^{2} + \gamma_{ij}\left(dx^{i}+\beta^{i}dt\right)\left(dx^{j}+\beta^{j}dt\right),
\end{equation}
where $\alpha$, $\beta^{i}$, and $\gamma_{ij}$ are the lapse, shift vector, and spatial metric, respectively, and where the indices $i, j, \ldots$ run over all $D-1$ spatial coordinates.  In spherical symmetry (which is preserved by the coordinate transformation (\ref{deftbarheightfunc})) the only non-vanishing component of the shift is the radial component $\beta^R$.  Comparing terms in the line elements (\ref{STlecompactnotnew}) and (\ref{dminus1plus1le}), we can identify the lapse as
\begin{equation}
\label{lapsesquh}
\alpha^{2} = \frac{f_{0}}{1-f_{0}^{2}h'^{2}},
\end{equation}
the shift as
\begin{equation}
\label{shifth}
\beta^{R} = \frac{f_{0}^{2}h'}{1-f_{0}^{2}h'^{2}},
\end{equation}
and the spatial metric as
\begin{equation}
\label{gammaijmatrix}
\gamma_{ij=}\left(
\begin{array}{cccccc}
\alpha^{-2}& 0 & 0 & \ldots & \ldots & 0\\
0 & R^{2} & 0 & \ldots & \ldots & 0\\
0 & 0 & R^{2}\sin^{2}\chi_{2} & 0 & \ldots & 0 \\
\vdots & \vdots & 0 & \ddots & 0 & 0\\
\vdots & \vdots & \vdots & 0 & \ddots & 0\\
0 & 0 & 0 & 0 & 0 & R^{2}\prod_{l=2}^{D-2}\sin^{2}\chi_{l}
\end{array}
\right).
\end{equation}
The determinant $\gamma$ of the spatial metric is
\begin{equation}
\label{defSTheightgamma}
\gamma = \alpha^{-2}R^{2\left(D-2\right)}\prod_{l=2}^{D-2}\left(\sin^{2}\chi_{l}\right)^{D-1-l}.
\end{equation}

We define the extrinsic curvature $K_{ij}$ so that 
\begin{equation}
\label{defKij}
\partial_{t}\gamma_{ij} = -2\alpha K_{ij} + D_{i}\beta_{j} + D_{j}\beta_{i},
\end{equation}
where $D_{i}$ is the covariant derivative operator associated with the spatial metric.  For time-independent spatial metrics like (\ref{gammaijmatrix}), the left-hand side of (\ref{defKij}) vanishes and we obtain
\begin{equation}
\label{defKijtimeind}
K_{ij} = \frac{1}{2\alpha}\left(D_{i}\beta_{j} + D_{j}\beta_{i}\right).
\end{equation}
Taking the trace of this equation we find that the mean curvature is given by
\begin{equation}
\label{defKsph}
K = \frac{1}{\alpha\gamma^{1/2}}\frac{d}{dR}\left(\gamma^{1/2}\beta^{R}\right).
\end{equation}
This equation is our starting point for imposing maximal slicing in Section \ref{MaxSlicSection} and $1+\rm{log}$ slicing in Section \ref{1pluslogSection}.

%==================================================================

\section{Maximal Slicing}
\label{MaxSlicSection}

Maximal slicing is defined  by requiring that the mean curvature vanish,
\begin{equation}
K = 0.
\end{equation}
Results for maximal slices of the Schwarzschild-Tangherlini spacetime have already been presented in \cite{NakAYS09}; the details included here are for the sake of completeness and reference in later sections.  In Section \ref{maxDgeq4subsect}, we derive a family of time-independent maximal slices of the Schwarzschild-Tangherlini spacetime for a general number of spacetime dimensions $D \geq 4$, and we specialize to $D=4$ and $D=5$ in Sections \ref{maxDeq4subsect} and \ref{maxDeq5subsect}, respectively.

%==================================================================

\subsection{General treatment}
\label{maxDgeq4subsect}

For maximal slicing, equation (\ref{defKsph}) reduces to
\begin{equation}
\label{timeindmaxslicingcondsph}
\frac{d}{dR}\left(\frac{R^{D-2}\beta^{R}}{\alpha}\right) = 0.
\end{equation}
Eliminating $\alpha$ and $\beta^R$ with the help of equations (\ref{lapsesquh}) and (\ref{shifth})  we obtain a first integral
\begin{equation}
\label{firstintegral}
R^{D-2}\left(\frac{f_{0}}{1-f_{0}^{2}h'^{2}}\right)^{1/2}f_{0}h' = C,
\end{equation}
where $C$ is a constant of integration.
It is convenient to write (\ref{firstintegral}) as
\begin{equation}
\label{firstintegralsubform}
f_{0}^{2}h'^{2} = \frac{C^{2}}{f_{0}R^{2\left(D-2\right)}+C^{2}},
\end{equation}
which we can then substitute back into equations (\ref{lapsesquh}) and (\ref{shifth}) to find the lapse
\begin{equation}
\label{alphaf}
\alpha = f\left(R;C\right),
\end{equation}
the shift
\begin{equation}
\label{shiftf}
\beta^{R} = \frac{Cf\left(R;C\right)}{R^{D-2}},
\end{equation}
and the spatial line element
\begin{equation}
\label{spatiallef}
dl^{2} = f^{-2}\left(R;C\right)dR^{2} + R^{2}d\Omega^{2}_{D-2}.
\end{equation}
Here the function $f(R;C)$ is given by
\begin{equation}
\label{deff}
f\left(R;C\right) = \left(1-\frac{\mu}{R^{D-3}}+\frac{C^{2}}{R^{2\left(D-2\right)}}\right)^{1/2}.
\end{equation}
The one-parameter family of spherically-symmetric, time-independent maximal slices of the Schwarzschild-Tangherlini spacetime is now parameterized  by the constant $C$.  For sufficiently small $C$ the slices end at a radius $R_0$ at which the lapse vanishes; from equation (\ref{alphaf}), this location is given by the largest root (see \cite{NakAYS09}) of the equation
\begin{equation} \label{R_0_max_slicing}
R_0^{2(D-2)} - \mu R_0^{D-1} + C^2 = 0.
\end{equation}

Two particular members of this family deserve special mention.  For $C=0$ the height function $h$ must be constant, the spacetime is therefore sliced by slices of constant Schwarschild-Tangherlini time $t$, and we recover the metric (\ref{STle}).  The other member of the family that we will be interested in is that for which the slice ends with a double-root of the squared 
lapse (\ref{alphaf}).  As we will show below, this choice singles out a trumpet slice.   In Section \ref{NumWorkSection} we will see that these slices act as ``attractors" in dynamical moving-puncture simulations.

Setting both $\alpha^2$ and the first derivative $\partial(\alpha^2)/\partial R$ to zero, we find 
that this double root occurs at the radius
\begin{equation}
\label{maxslicingthroat}
\tilde R_{0}=\left(\frac{\mu\left(D-1\right)}{2\left(D-2\right)}\right)^{1/\left(D-3\right)}
\end{equation}
for the value of $C$ given by
\begin{equation}
\label{Csquspecial}
\tilde{C}^{2}\equiv\left(\frac{D-3}{D-1}\right)\left(\frac{\mu\left(D-1\right)}{2\left(D-2\right)}\right)^{2\left(D-2\right)/\left(D-3\right)},
\end{equation}
where the tilde denotes the special value for the trumpet slice
(see also \cite{NakAYS09}).
 
Before closing this section we analyze the asymptotic properties of these slices in a neighborhood of $R_0$.   Given the spatial metric (\ref{gammaijmatrix}), we can compute the proper distance $\Delta$ between a point on the limit surface, $(R_0,\chi_i)$, and a point at $(R_1,\chi_i)$ from the integral
\begin{equation} \label{prop_dist}
\Delta = \int_{R_0}^{R_1} \frac{dR}{\alpha}. 
\end{equation}
By definition, the lapse $\alpha$ vanishes at $R_0$.   Whether or not this integral is finite therefore depends on the behavior of $\alpha$ in the neighborhood of $R_0$.  From its definition in Eqs.~(\ref{alphaf}) and (\ref{deff}), we see that the square of the lapse can be expanded as
\begin{equation}
	\alpha^2 = A_1(R - R_0) + A_2 (R - R_0)^2 + \cdots \ .
\end{equation}
For generic values of $C$, $R = R_0$ is a single root of $\alpha^2$ and hence $A_1$ is nonzero. Therefore the lapse behaves like $\alpha \sim (R - R_0)^{1/2}$ near $R_0$ and the integral in Eq.~(\ref{prop_dist}) is finite. Thus, for generic $C$, the proper distance to the limiting surface at $R_0$ is finite. 

For the special slice $C = \tilde C$, on the other hand, $R = \tilde R_0$ is a double root of $\alpha^2$ and therefore $A_1$ vanishes. The lapse then behaves like  $\alpha \sim (R - \tilde R_0)$ near $\tilde R_0$ and the integral in Eq.~(\ref{prop_dist}) diverges. The proper distance to the limiting surface at $\tilde R_0$ is infinite, even though the areal radius is non-zero and finite (see \cite{NakAYS09} for an alternative derivation of this property).  An embedding diagram of the slice would result in a figure similar to that in Fig.~2 of \cite{HanHOBO08}; given the appearance in an embedding diagram these slices are referred to as ``trumpet" slices.

As in Section \ref{SchwTanSection}, it would be useful to transform these slices to isotropic coordinates.  Instead of equation (\ref{defrintegral}), we now have
\begin{equation}
\label{defrintegralCne0}
\pm\int\frac{dr}{r} = \int\frac{1}{f}\frac{dR}{R} = 
\int\frac{R^{D-3}dR}{\sqrt{R^{2\left(D-2\right)}-\mu R^{D-1}+ C^{2}}}.
\end{equation}
For $C = \tilde C$, the expression under the square root on the right hand side has a double root at $R = \tilde R_0$, which simplifies the integral for both $D=4$ and $D=5$.

%=====================================================================

\subsection{Four-dimensional spacetimes}
\label{maxDeq4subsect}

For four-dimensional spacetimes, $D = 4$, we recover the well-known family of time-independent maximal slices \cite{EstWCDST,Rei73,BeiO98,HanHOBO08,BauS10}.  In particular, with $\Omega_{2}=4\pi$ and $\mu=2GM$,  the lapse becomes
\begin{equation}
\label{fmaxdequals4}
\alpha=f\left(R;C\right) = \left(1-\frac{2GM}{R}+\frac{C^{2}}{R^{4}}\right)^{1/2}.
\end{equation}
With the lapse known, the shift, spatial metric, and extrinsic curvature can be calculated from (\ref{shiftf}), (\ref{gammaijmatrix}), and (\ref{defKijtimeind}), respectively.  Also, for the special value of $\tilde C$, which now becomes
\begin{equation}
\tilde C = \frac{3 \sqrt{3} (GM)^2}{4},
\end{equation}
the transformation to isotropic coordinates can be carried out by integrating equation (\ref{defrintegralCne0}) (see \cite{BauN07}). 

%=====================================================================

\subsection{Five-dimensional spacetimes}
\label{maxDeq5subsect}

For five-dimensional spacetimes, $D=5$, we recover the results of \cite{NakAYS09}.  We note, however, a difference in notation: our $R$ is their $r$ and vice versa.  With $\Omega_{3}=2\pi^{2}$ and $\mu=8GM/(3\pi)$ we now find the lapse
\begin{equation}
\label{fmaxdequals5}
\alpha=f\left(R;C\right)=\left(1-\frac{8GM}{3\pi R^{2}}+\frac{C^{2}}{R^6}\right)^{1/2}.
\end{equation}
With the lapse known, the shift, spatial metric, and extrinsic curvature can again be calculated from (\ref{shiftf}), (\ref{gammaijmatrix}), and (\ref{defKijtimeind}), respectively.   The trumpet geometry is realized for 
\begin{equation}
\tilde C = \frac{2^{11/2}}{3^3}\left(\frac{GM}{\pi}\right)^{3/2},
\end{equation}
and its limiting surface is at
\begin{equation} \label{R_0_tilde_5D}
\tilde R_0 = \frac{4}{3} \sqrt{\frac{GM}{\pi}}.
\end{equation}
For the trumpet slice, equation (\ref{defrintegralCne0}) can again be integrated analytically (see \cite{NakAYS09}) to obtain the solution in isotropic coordinates.

%=====================================================================

\section{Stationary 1 + log Slicing}
\label{1pluslogSection}

We now turn to stationary 1+log slicing
\begin{equation}
\label{timeind1pluslogslicingcond}
\beta^{i}\partial_{i}\alpha = n\alpha K,
\end{equation}
where $n$ is a constant.   The most common choice for $n$ is $n=2$, but, following \cite{HanHOBO08}, we will pursue a more general treatment.  In particular, we will recover the maximal slicing results of Section \ref{MaxSlicSection} in the limit $n \rightarrow \infty$.  Mirroring Section \ref{MaxSlicSection}, we derive a family of time-independent $1+\rm{log}$ slices of the Schwarzschild-Tangherlini spacetime for arbitrary $D \geq 4$ in Section \ref{1pluslogDgeq4subsect}, and specialize to $D=4$ and $D=5$ in Sections \ref{1pluslogDeq4subsect} and \ref{1pluslogDeq5subsect}.

%=====================================================================

\subsection{General treatment}
\label{1pluslogDgeq4subsect}

Using equations (\ref{STlecompactnotnew}), (\ref{dminus1plus1le}), and (\ref{defKsph}), we can rewrite (\ref{timeind1pluslogslicingcond}) as
\begin{equation}
\label{timeind1pluslogslicingcondsph}
\frac{d}{dR}\alpha = \frac{n}{\gamma^{1/2}\beta^{R}} \frac{d}{dR} \left(\gamma^{1/2}\beta^{R}\right).
\end{equation}
It is convenient to use (\ref{lapsesquh}) and (\ref{shifth}) to find
\begin{equation}
\label{shift1pluslog}
\beta^{R} = \alpha\sqrt{\alpha^{2}-f_{0}},
\end{equation}
and to eliminate $\beta^{R}$ in (\ref{timeind1pluslogslicingcondsph}). This substitution results in
\begin{equation}
\label{lapseODE1pluslog}
\frac{d}{dR} \alpha = n\frac{d}{dR} \ln \left(R^{\left(D-2\right)}\sqrt{\alpha^{2}-f_{0}}\right),
\end{equation}
which can be integrated immediately to yield
\begin{equation}
\label{lapsesolvedbyint}
\alpha = \mbox{const} +
	n\ln{\left(R^{\left(D-2\right)}\sqrt{\alpha^{2}-f_{0}}\right)},
\end{equation}
or equivalently,
\begin{equation}
\label{lapsesqunice}
\alpha^{2} = f_{0} + 
	\frac{C^{2}\left(n\right)e^{2\alpha/n}}{R^{2\left(D-2\right)}}.
\end{equation}
Here $C$ is again a constant of integration, but we point out that it now depends on the constant $n$.
Equation (\ref{lapsesqunice}) is a transcendental equation for the lapse $\alpha$, and solutions, when they exist, can be found numerically.  With the solution for the lapse in hand, the shift, spatial metric, and extrinsic curvature can be found from (\ref{shift1pluslog}), (\ref{gammaijmatrix}), and (\ref{defKijtimeind}).

As for maximal slicing, some values of $C^{2}\left(n\right)$ are special.  Setting $C^{2}\left(n\right)=0$ corresponds to the usual $t=\rm{const}$ slices, recovering the metric (\ref{STle}).  Another solution of interest is obtained by requiring equation (\ref{lapseODE1pluslog}) to be regular for all $\alpha\geq 0$.   Following the method given in \cite{HanHOBO08} for $D=4$, we can use the regularity condition to determine the corresponding values $\tilde C^{2}\left(n\right)$ for $D\geq 4$.  We begin by rewriting (\ref{lapseODE1pluslog}) as
\begin{equation}
\label{lapseODE1pluslogalt}
\frac{d}{dR} \alpha = -\frac{n\left(\left(D-2\right)R^{-1}\left(\alpha^{2}-f_{0}\right)-\frac{1}{2}f_{0}'\right)}{f_{0}+n\alpha-\alpha^{2}}.
\end{equation}
If we require regularity when $\alpha\geq 0$, the numerator and denominator must vanish at the same value of $R$.  The numerator is zero when
\begin{equation}
\label{lapseequnumcond}
\alpha = \sqrt{f_{0}+\frac{Rf_{0}'}{2\left(D-2\right)}}.
\end{equation}
Substituting this value for the lapse and $f_{0}$ from (\ref{deff0}) into the right hand side of (\ref{lapseODE1pluslogalt}), we see that its denominator vanishes when
\begin{equation}
\label{lapseequdenomcond}
R^{2D-6}+\frac{\left(1-D\right)\mu}{2\left(D-2\right)}R^{D-3}-\frac{\left(D-3\right)^{2}\mu^{2}}{4n^{2}\left(D-2\right)^{2}}=0.
\end{equation}
This equation is quadratic in $R^{D-3}$, and so it is easily solved for the critical value $R_c$ for which the numerator and denominator of (\ref{lapseODE1pluslogalt}) vanish simultaneously.  The positive real root is
\begin{equation}
\label{Rcfound}
%R_{c} = \left(\frac{ \left(D-1\right)n^{2}\mu/2 +\mu\sqrt{\left(D-3\right)^{2}n^{2}+ \left(D-1\right)^{2}n^{4}/4}}{2\left(D-2\right)n^{2}}\right)^{1/(D-3)}.
\begin{array}{rcl}
&&\displaystyle R_{c} = \\ 
&&\displaystyle \left(\frac{ \left(D-1\right)\mu +\mu\sqrt{4\left(D-3\right)^{2}n^{-2}+ \left(D-1\right)^{2}}}{4\left(D-2\right)}\right)^{\frac{1}{D-3}}.
\end{array}
\end{equation}
The lapse at $R_{c}$ can be calculated from (\ref{lapseequnumcond}) to be
\begin{equation}
\label{alphac}
\alpha_{c} = \sqrt{\frac{\sqrt{4\left(D-3\right)^{2}+\left(D-1\right)^{2}n^{2}}-\left(D-1\right)n}{\sqrt{4\left(D-3\right)^{2}+\left(D-1\right)^{2}n^{2}}+\left(D-1\right)n}}.
\end{equation}
Inserting equations (\ref{Rcfound}) and (\ref{alphac}) into (\ref{lapsesqunice}) we now find $\tilde{C}^{2}\left(n\right)$, defined as the special value of $C^{2}\left(n\right)$ that makes (\ref{lapseODE1pluslogalt}) regular,
\begin{equation}
%\begin{eqnarray}
\label{Csqspecial1pluslog}
%\tilde{C}^{2}\left(n\right) &=& \frac{D-3}{2^{\left(3D-5\right)/\left(D-3\right)}}\left(\frac{\mu}{D-2}\right)^{\left(2\left(D-2\right)\right)/\left(D-3\right)}\left(e^{-\frac{2\alpha_{c}}{n}}\right)\times\nonumber\\&&\left(D-1+\frac{\sqrt{4\left(D-3\right)^{2}+\left(D-1\right)^{2}n^{2}}}{n}\right)^{\left(D-1\right)/\left(D-3\right)}.
%\end{eqnarray}
\begin{array}{rcl}
&&\displaystyle \tilde{C}^{2}\left(n\right) = \frac{D-3}{2^{\frac{3D-5}{D-3}}}\left(\frac{\mu}{D-2}\right)^{\frac{2\left(D-2\right)}{D-3}}\left(e^{-\frac{2\alpha_{c}}{n}}\right)\\
&&\displaystyle \times\left(D-1+\frac{\sqrt{4\left(D-3\right)^{2}+\left(D-1\right)^{2}n^{2}}}{n}\right)^{\frac{D-1}{D-3}}.
\end{array}
\end{equation}

\begin{figure}
\begin{center}
\includegraphics[width=3.25in]{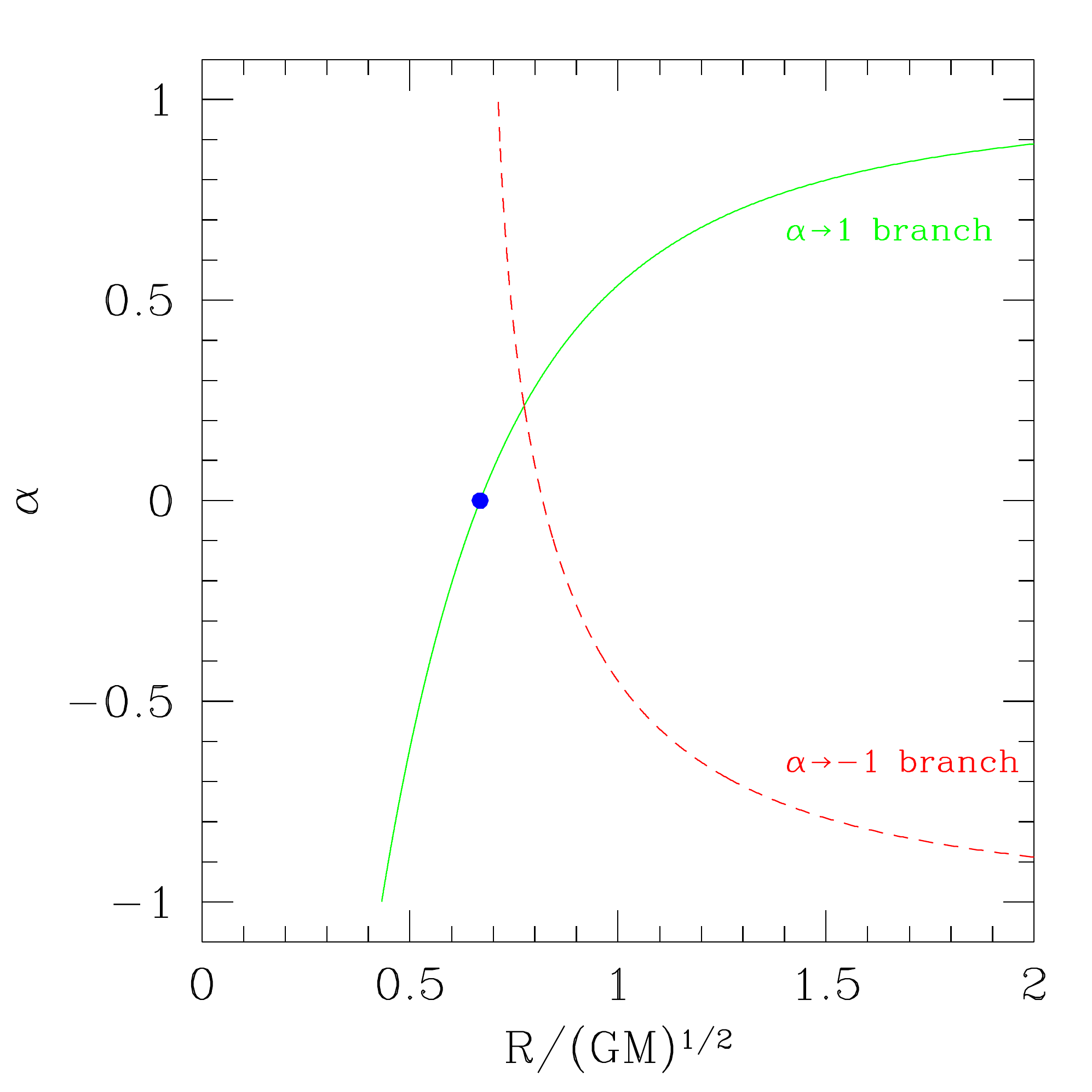}
\end{center}
\caption{The two branches of the lapse $\alpha$ as a function of the radius $R$ for $D=5$, $n=2$, and $C = \tilde C$.  We are interested in the ``positive" branch that approaches $\alpha = 1$ at infinity.  This branch has a root (with $\alpha = 0$) at a smaller value of $R$ than the ``negative" branch.}
\label{Fig:alpha_vs_R}
\end{figure}

For most of the work that follows we consider only the special slices with $C=\tilde C$.  The inner boundary of these slices is defined by the location at which the lapse goes to zero.  We will refer to this location as the throat, and will call its radius $\tilde R_0$.  To identify $\tilde R_0$, we first observe that, at least for sufficiently large $R$, equation (\ref{lapsesqunice}) admits two solutions for the lapse $\alpha$.  For $R \rightarrow \infty$, one of the branches of solutions approaches $\alpha = 1$, while the other approaches $\alpha = -1$.  This behavior is displayed in Fig.~\ref{Fig:alpha_vs_R}, where we graph $\alpha$ versus $R$ for $D=5$ and $n=2$, for the special value $C = \tilde C$.  We are interested in the ``positive" branch of solutions, which approaches $\alpha = 1$ asymptotically.  As can be seen from Fig.~\ref{Fig:alpha_vs_R}, this positive branch has a root (with $\alpha = 0$) at a smaller value of $R$ than the negative branch.  In general, then, we define the throat as the location $\tilde R_0$ at which the positive branch of solutions for $\alpha$ vanishes.  For $D=4$ and $D=5$, this corresponds to the {\em smallest} root of equation (\ref{lapsesqunice}) with $\alpha = 0$,
\begin{equation}
\label{R0poly}
\tilde R_0^{2\left(D-2\right)}-\mu \tilde R_0^{D-1} + \tilde C^{2}\left(n\right) = 0
\end{equation}
(see also the discussion in \cite{HanHOBO08}).  In the limit $n\rightarrow\infty$, the critical value of the lapse $\alpha_{c}\rightarrow 0$ and the slice ends in a double root, so this is still consistent with choosing the {\em largest} root in the case of maximal slicing.  We will discuss explicit solutions for $\tilde R_{0}\left(n\right)$ for $D=4$ and $D=5$ in Sections \ref{1pluslogDeq4subsect} and \ref{1pluslogDeq5subsect} below.

We can demonstrate that the special slice with $C = \tilde C$ is a trumpet slice.  To do so, we assume that close to $\tilde R_0$, 
$\alpha$ can be expanded as a power series 
\begin{equation}
	\alpha = \left(\frac{\delta}{\tilde R_0}\right)^m \left( \alpha_0 + \alpha_1 \frac{\delta}{\tilde R_0} \cdots \right) 
\end{equation}
where $R = \tilde R_0 + \delta$. 
Since $\alpha = 0$ for $\delta = 0$ we must have $m>0$. 
Inserting this into equation (\ref{lapsesqunice}), with $C = \tilde C$, 
and expanding terms to first leading order, we obtain 
\begin{equation} \label{alpha_expansion3}
\begin{array}{rcl}
& & \displaystyle \alpha_0^2 \left( \frac{\delta}{\tilde R_0} \right)^{2m} =  \frac{\tilde C^2}{\tilde R_0^{2(D-2)}} \,\frac{2}{n} \left( \frac{\delta}{\tilde R_0} \right)^m \alpha_0\\ 
& & ~~~~ \displaystyle + \left( (D-3) \frac{\mu}{\tilde R_0^{D-3}} + 2(2 - D) \frac{\tilde C^2}{\tilde R_0^{2(D-2)}} \right) \frac{\delta}{\tilde R_0}. 
\end{array}
\end{equation}
Matching exponents of $\delta$ we identify $m = 1$, so that $\alpha \sim \delta = R-\tilde R_0$ near $R = \tilde R_0$. Therefore the proper distance from any point outside the throat to the throat at $\tilde R_0$, given by the integral (\ref{prop_dist}), diverges.  As before, this demonstrates that this slice features a trumpet geometry.

%For maximal slicing we found this diverging behavior only for the special slices with $C = \tilde C$, while our analysis suggests that all $1+\rm{log}$ slices are trumpet slices.  This is not correct, however, since only those slices with $C = \tilde C$ remain regular for all $\alpha = 0$, so that we should apply our analysis above only to these special slices.

In the limit $n\rightarrow\infty$, the $1+\rm{log}$ slicing condition (\ref{timeind1pluslogslicingcond}) approaches the maximal slicing condition $K=0$, so we should recover the maximal slicing results from Section \ref{MaxSlicSection}.  From (\ref{lapsesqunice}) we have
\begin{equation}
\label{alphasqumaxlimit}
\alpha^{2} = f_{0} + \frac{C^{2}}{R^{2\left(D-2\right)}},
\end{equation}
consistent with (\ref{alphaf}).  Similarly, taking the limit $n \rightarrow \infty$ in equation (\ref{Csqspecial1pluslog}) we recover (\ref{Csquspecial}).  Finally, letting $n\rightarrow\infty$ in (\ref{Rcfound}) results in
\begin{equation}
\label{Rcmaxlimit}
R_{c} = \left(\frac{\mu\left(D-1\right)}{2\left(D-2\right)}\right)^{1/\left(D-3\right)}.
\end{equation}
Substitution into (\ref{R0poly}) shows that this coincides with the location of the throat, so that $R_c = \tilde R_0$ in this limit, meaning that we have recovered $\tilde R_0$ in the maximal slicing limit (\ref{maxslicingthroat}).

As in Section \ref{SchwTanSection}, it would useful to transform these slices to isotropic coordinates.   While it is impossible to carry out the necessary integrations analytically,
given the transcendental nature of equation (\ref{lapsesqunice}), the equations can be integrated numerically using the techniques presented in \cite{HanHOBO08}.

%======================================================================

\subsection{Four-dimensional spacetimes}
\label{1pluslogDeq4subsect}

\begin{figure}
\begin{center}
\includegraphics[width=3.25in]{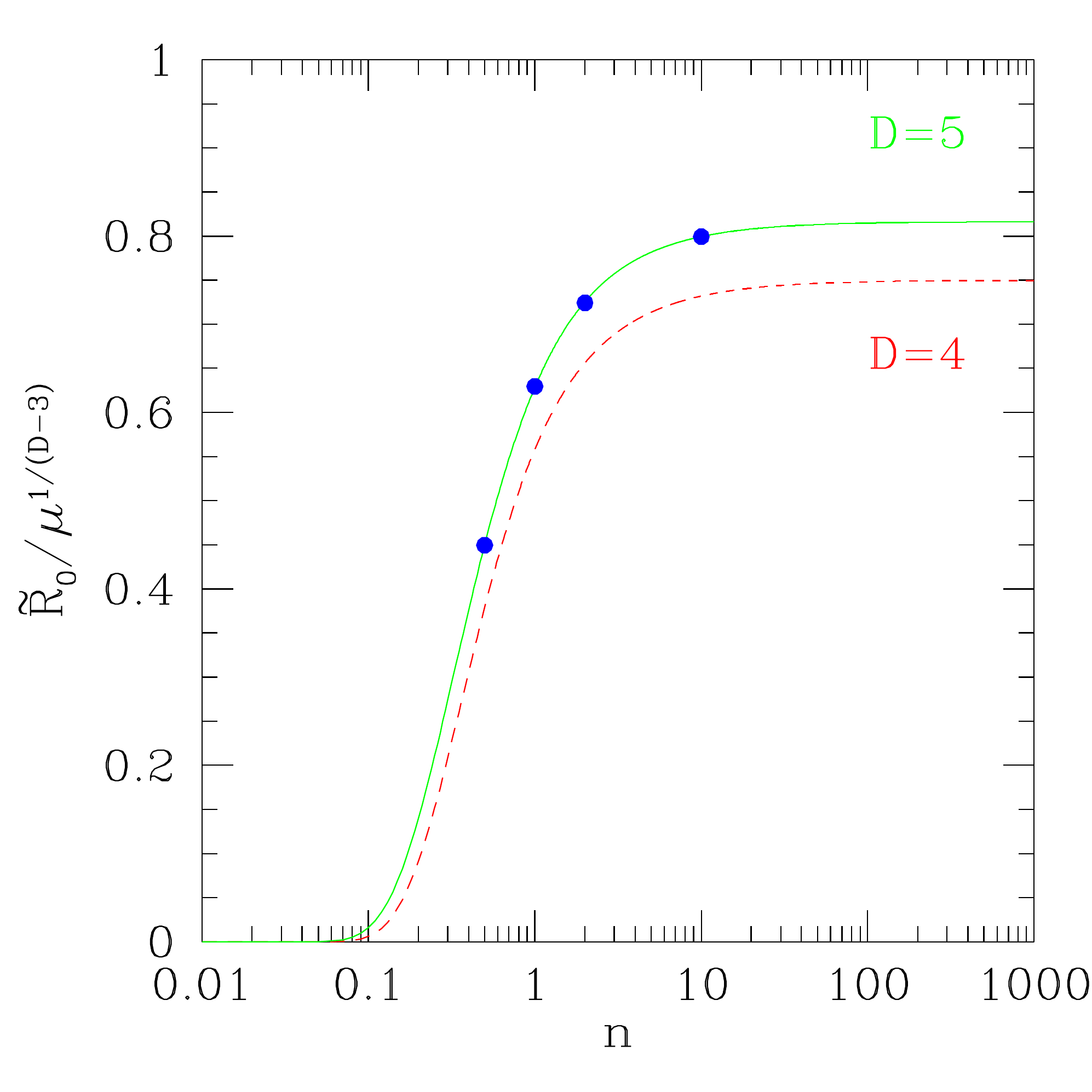}
\caption{The throat location $\tilde R_{0}$ as a function of the coefficient $n$ in the $1+\rm{log}$ slicing condition (\ref{timeind1pluslogslicingcond}) for $D=4$ (dashed) and $D=5$ (solid).  The points are the corresponding values found by our numerical simulations described in Section \ref{NumWorkSection}.  Specifically, we extrapolated the lapse $\alpha$ linearly to zero to find the throat location $\tilde R_0$ for any grid resolution, and then used Richardson extrapolation using our two highest-order resolution results to obtain the values shown here.}
\label{fig:R_0dimlessvsn}
\end{center}
\end{figure}

For four-dimensional spacetimes, the above results reduce to those of \cite{HanHOBO08}.  In particular, for $D=4$, equation (\ref{R0poly}) becomes a quartic  equation
\begin{equation}
\label{R0polyD4}
\tilde R_0^{4}-\mu \tilde R_0^{3}+\tilde{C}^{2}\left(n\right)=0.
\end{equation}
The real root consistent with the positive branch of the lapse is
\begin{equation}
\label{R0polydequals4solved}
\tilde R_{0}\left(n\right)=\frac{\mu}{2}\left(\frac{1}{2}+\sqrt{\frac{1}{4}+Z}-\sqrt{\frac{1}{2}-Z+\frac{1}{4\sqrt{1/4+Z}}}\right),
\end{equation}
where
\begin{equation}
\label{defZdequals4}
Z=\frac{4\left(\frac{2}{3}\right)^{1/3}\tilde{C}^{2}\left(n\right)}{Y\mu^{4}}+\frac{Y}{2^{1/3}3^{2/3}},
\end{equation}
and
\begin{equation}
\label{defYdequals4}
Y=\left(\left(9+\sqrt{81-768\tilde{C}^{2}\left(n\right)\mu^{-4}}\right)\mu^{-4}\tilde{C}^{2}\left(n\right)\right)^{1/3}.
\end{equation}  
We graph these solutions in Fig.~\ref{fig:R_0dimlessvsn} (compare with Fig.~3 of 
\cite{HanHOBO08}, where these solutions are given only graphically).

For comparison with \cite{HanHOBO08}, we note that (\ref{lapseODE1pluslogalt}) can be written as
\begin{equation}
\label{lapseODE1pluslogaltD4}
\begin{array}{rcl}
\displaystyle \frac{d}{dR}\alpha & = & \displaystyle -\frac{n}{R\left(R^{D-3}\left(f_{0}+n\alpha-\alpha^{2}\right)\right)}\\[3mm]
& & \displaystyle \times \Big( -\left(D-2\right)R^{D-3}f_{0}-\frac{1}{2}R^{D-2}f_{0}'\\
& & ~~~~\displaystyle +\left(D-2\right)R^{D-3}\alpha^{2} \Big),
\end{array}
\end{equation}
which, for $D=4$, reduces to
\begin{equation}
\label{lapseODE1pluslogaltD4Hcheck}
\frac{d}{dR}\alpha=-\frac{n\left(3GM-2R+2R\alpha^{2}\right)}{R\left(R-2GM+nR\alpha-R\alpha^{2}\right)}.
\end{equation}
From this equation, the remaining results of Section II~D~1 of \cite{HanHOBO08} can be derived.

%====================================================================

\subsection{Five-dimensional spacetimes}
\label{1pluslogDeq5subsect}

For $D=5$, equation (\ref{R0poly}) becomes a cubic equation for the $\tilde R_0^2(n)$,
\begin{equation}
\label{R0polydequals5}
\tilde R_{0}^{6}-\mu \tilde R_{0}^{4}+\tilde{C}^{2}\left(n\right)=0,
\end{equation}
which can again be solved exactly.  The root that is real, positive, and consistent with the positive branch of the lapse is
\begin{equation}
\label{R01pluslogD5}
\begin{array}{rcl}
& & \displaystyle \tilde R_{0} = \\
& & \displaystyle \frac{\mu^{1/2}}{3^{1/2}} \sqrt{ 2\cos \left( \frac{1}{3} \arccos \left(1- \frac{27\tilde C^{2}\left(n\right)}{2\mu^{3}} \right) + \frac{4\pi}{3}  \right)+1 } .
\end{array}
\end{equation}
We show a graph of $\tilde R_0$ in Fig.~\ref{fig:R_0dimlessvsn}.
 
With $\Omega_{3}=2\pi^{2}$, $\mu=8GM/(3\pi)$, and $f_{0}=1- 8GM/(3\pi R^{2})$, the solution (\ref{lapsesqunice}) of the stationary 1+log slicing condition becomes
% equation (\ref{lapseODE1pluslogalt}) becomes
%\begin{equation}
%\frac{d}{dR}\alpha=-\frac{n\left(16GM/\left(3\pi\right)-3R^{2}+3R^{2}\alpha^{2}\right)}{R\left(R^{2}-8GM/\left(3\pi\right)+nR^{2}\alpha-R^{2}\alpha^{2}\right)}.
%\end{equation}
\begin{equation}
\alpha^{2}=1-\frac{8GM}{3\pi R^{2}}+\frac{C^{2}\left(n\right)e^{2\alpha/n}}{R^{6}}.
\end{equation}
This determines $\alpha$ as a function of $R$.  When solutions to this equation exist, they can be found numerically with a root-finding routine by starting with an appropriate guess on the positive branch of the lapse.  With the lapse known, the shift, spatial metric, and extrinsic curvature can be calculated from (\ref{shift1pluslog}), (\ref{gammaijmatrix}), and (\ref{defKijtimeind}), respectively.  For some purposes, it is sufficient to find $R\left(\alpha\right)$ instead of $\alpha\left(R\right)$.  The real, positive solutions of relevance here are closely related to the solution (\ref{R01pluslogD5}) for $\tilde R_{0}\left(n\right)$:
\begin{equation}
\label{R1ofalpha}
\begin{array}{rcl}
& & \displaystyle R_{1}\left(\alpha\right)= \sqrt{\frac{\mu}{3\left(1-\alpha^{2}\right)}}\\
& & \displaystyle \times\sqrt{ 2\cos \left( \frac{1}{3} \arccos \left(1- \frac{27A}{2} \right) + \frac{4\pi}{3}  \right)+1 },
\end{array}
\end{equation}
and
\begin{equation}
\label{R2ofalpha}
\begin{array}{rcl}
& & \displaystyle R_{2}\left(\alpha\right)= \sqrt{\frac{\mu}{3\left(1-\alpha^{2}\right)}}\\
& & \displaystyle \times\sqrt{ 2\cos \left( \frac{1}{3} \arccos \left(1- \frac{27A}{2} \right)\right)+1 },
\end{array}
\end{equation}
where
\begin{equation}
A=\frac{\left(1-\alpha^{2}\right)^{2}}{\mu^{3}}C^{2}\left(n\right)e^{2\alpha/n}.
\end{equation}
With $C=\tilde C$, the positive branch of the lapse is given by $R_{1}\left(\alpha\right)$ for $0\leq\alpha\leq \alpha_{c}$ and $R_{2}\left(\alpha\right)$ for $\alpha_{c}\leq\alpha\leq 1$.  Solutions for $D=4$ can be constructed similarly from solutions to equation (\ref{R0polyD4}).
  
The critical radius (\ref{Rcfound}) is now 
\begin{equation}
R_{c}=\frac{2}{3}\left(\frac{2GM}{\pi n}\left(n+\sqrt{1+n^{2}}\right)\right)^{1/2},
\end{equation}
and the lapse (\ref{alphac}) at $R = R_c$ is 
\begin{equation}
\alpha_{c}=\sqrt{\frac{\sqrt{1+n^{2}}-n}{\sqrt{1+n^{2}}+n}}.
\end{equation}
Finally, we find $\tilde C$ from equation (\ref{Csqspecial1pluslog}), 
\begin{equation}
\tilde{C}^{2}\left(n\right)=\frac{2^{9}\left(GM\right)^{3}}{3^{6}\pi^{3}}\left(1+\frac{1}{n}\sqrt{1+n^{2}}\right)^{2}e^{-2\alpha_{c}/n}.
\end{equation}
As before we can recover the maximal slicing results of Section \ref{maxDeq5subsect} by letting $n \rightarrow \infty$.  From equations (\ref{Rcmaxlimit}) and (\ref{R01pluslogD5}) we then obtain
\begin{equation}
R_{c}=\tilde R_{0}=\frac{4}{3}\sqrt{\frac{GM}{\pi}},
\end{equation}
which agrees with equation (\ref{R_0_tilde_5D}).

Most numerical relativity simulations adopt $n=2$.  In this case equation (\ref{Rcfound}) yields
\begin{equation}
\label{Rcnequals2dequals5}
R_{c} = \frac{2}{3}\sqrt{\frac{2+\sqrt{5}}{\pi}}\sqrt{GM}\approx 0.774132\sqrt{GM},
\end{equation}
and from (\ref{Csqspecial1pluslog}) we have
\begin{eqnarray}
\label{Csqnequals2dequals5}
\tilde C^{2}\left(2\right)&=&\frac{128\left(2+\sqrt{5}\right)^{2}e^{-\sqrt{\left(\sqrt{5}-2\right)/\left(\sqrt{5}+2\right)}}}{729\pi^{3}} (GM)^{3}\nonumber\\
&\approx& 0.0802483 \, (GM)^{3}.
\end{eqnarray}
The throat (\ref{R01pluslogD5}) is located at 
\begin{equation}
\label{throatlocationnequals2dequals5}
\tilde R_{0}\approx 0.725474\mu^{\frac{1}{2}}\approx 0.668392\sqrt{GM}.
\end{equation}
For $D=5$, the event horizon in the Schwarzschild-Tangherlini spacetime is located at $R_{\rm EH} = \sqrt{8GM/(3\pi)} = 0.921318 \sqrt{GM}$.  At the horizon, the lapse is
\begin{equation}
\label{horizonlapsenequals2dequals5}
\alpha_{\rm EH} \approx 0.454702.
\end{equation}

%===================================================================

\section{Numerical Work}
\label{NumWorkSection}

In this section we describe dynamical numerical simulations of the Schwarzschild-Tangherlini spacetime for $D=5$.  More specifically, we adopt as initial data the $t=const$ slices of the Schwarzschild-Tangherlini spacetime in isotropic coordinates, as described in Section \ref{SchwTanSection}, and evolve these data with moving-puncture gauge conditions.  We will see that the trumpet solutions derived in the previous sections indeed act as ``attractors" in dynamical simulations, meaning that dynamical simulations settle down to these solutions at asymptotically late times, even when they start with very different initial data.

%=====================================================================

\subsection{Numerical Method}
\label{NumMethodSection}

Our numerical code is based on the third-order finite-difference code of \cite{Bro09}.  This code implements the BSSN equations \cite{ShiN95,BauS99} for $D=4$ in spherical symmetry, and imposes spherical symmetry with the help of the ``cartoon" method \cite{AlcBBHSTT01}.  Details of the implementation of this method can be found in Appendix A of \cite{Bro09}.  Several modifications had to be implemented to adopt this code to five spacetime dimensions.  Clearly, all indices of spatial tensors had to be extended to four spatial dimensions, and the interpolation for the cartoon method had to be adjusted for the extra dimension.  In addition, the finite differencing stencil was updated to account for the fourth spatial dimension.  Furthermore, even though Einstein's equations take the same form for all spacetime dimensions $D$, the BSSN equations for $D=5$ are different from the corresponding equations in $D=4$ (see, e.g.~\cite{YosS09}).  The reason for this is the use of tracefree tensors in the BSSN formalism.  In particular, consider the decomposition of the extrinsic curvature $K_{ij}$ into its trace $K$ and its traceless part $A_{ij}$,
\begin{equation}
A_{ij} = K_{ij} - \frac{1}{D-1} \gamma_{ij} K.
\end{equation}
Since the BSSN equations are formulated in terms of $A_{ij}$ and $K$ instead of $K_{ij}$, factors that depend on $D$ appear in several places.  Following \cite{YosS09}, we also use the variable $\chi \equiv \psi^{-2}$ instead of the conformal factor $\psi$ in our numerical simulations.

We tested features of our code that are specific to $D=5$ with the help of a number of test problems, including a spherically symmetric standing wave.  We also monitored constraint violations during our simulations and verified that they converged to zero at the expected rate.

%=====================================================================

\subsection{Numerical Results}
\label{NumResultsSubsection}

In all our dynamical simulations we start with data on a slice of constant Schwarzschild-Tangherlini time, for which the conformal factor is given by equation (\ref{ST_conf_factor}).   If these data were evolved with the Killing lapse and shift that can be identified from the metric (\ref{STle}), the metric would remain time-independent.  Instead, we set
\begin{equation} \label{alpha_beta_init}
\alpha = 1,~~~~~~~\beta^i=0
\end{equation}
initially (at $t=0$), and subsequently evolve the lapse and shift using moving puncture coordinate conditions \cite{CamLMZ06,BakCCKM06a}, namely the 1+log slicing condition for the lapse \cite{BonMSS95} and the $\Gamma$-driver condition \cite{AlcBDGHHHKPST05}.  We will consider two different flavors of the 1+log slicing condition, namely an ``advective" and a ``non-advective" version.

\subsubsection{1+log slicing}

The ``advective" 1+log slicing condition is given by 
\begin{equation} \label{1+log}
(\partial_t - \beta^i \partial_i) \alpha = - n \alpha K,
\end{equation}
where $n$ is again a constant.  At late times, when the solution settles down and becomes time-independent, this condition reduces to the stationary 1+log condition (\ref{timeind1pluslogslicingcond}).  

Our numerical results demonstrate that simulations with the advective 1+log slicing condition (\ref{1+log}) indeed settle down to the stationary 1+log results of Section \ref{1pluslogDeq5subsect}.  In the following we show results that were obtained with $N=5001$ uniform gridpoints (not counting the buffer points), with the outer boundary imposed at $r_{\rm max} = 50\sqrt{GM}$.  We also chose $n=2$ for all results shown in this section.  (We performed simulations for other values of $n$ and $N$ to obtain the results for the radius of the throat $\tilde R_0$ shown in Fig.~\ref{fig:R_0dimlessvsn}.)

\begin{figure} 
\begin{center}
\includegraphics[width=3.25in]{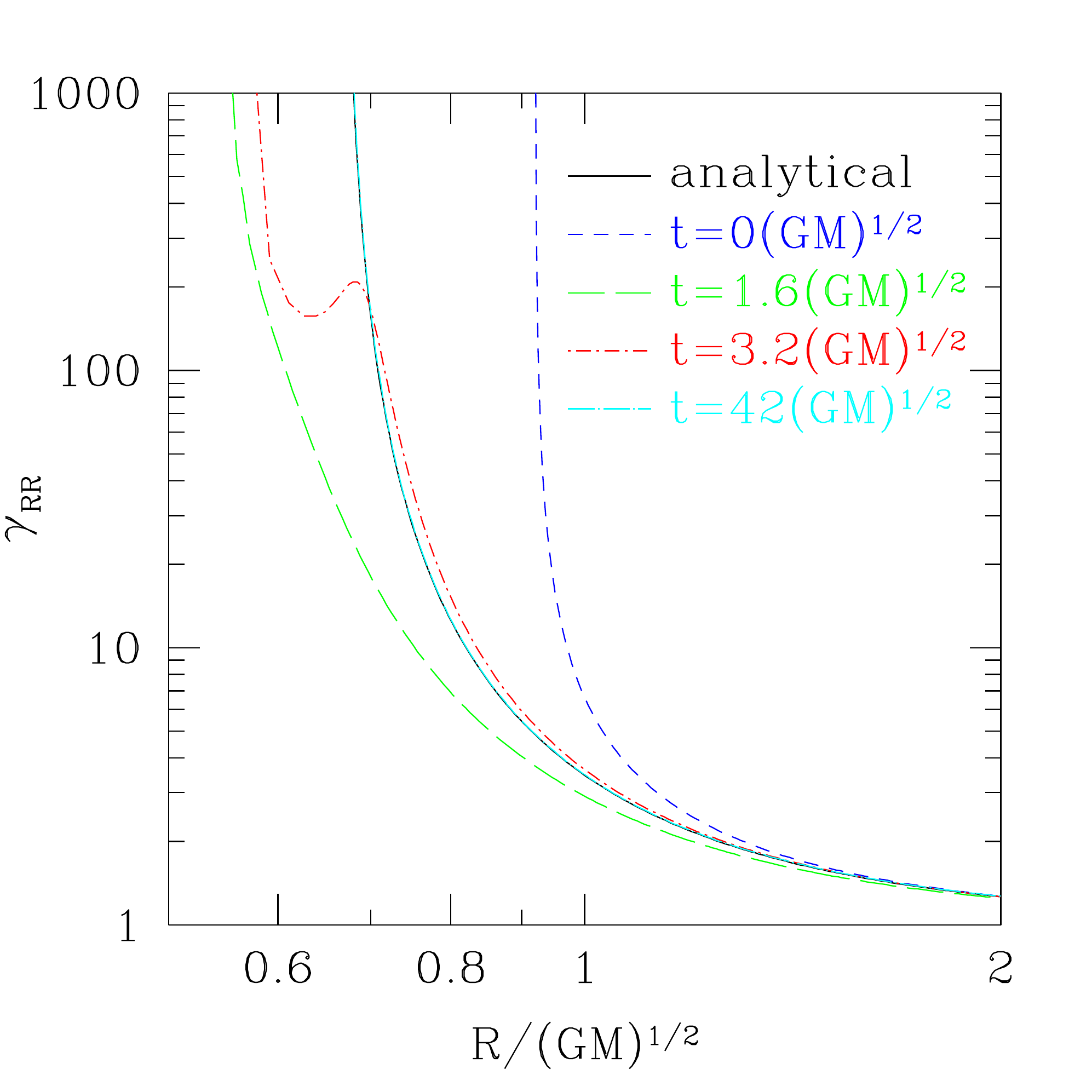}
\end{center}
\caption{The metric component $\gamma_{RR}$ as a function of the areal radius $R$ for a dynamical evolution of the $D=5$ Schwarzschild-Tangherlini spacetime with the advective 1+log slicing condition (\ref{1+log}) for $n=2$.  We show $\gamma_{RR}$ at a number of different times, starting with the initial data (\ref{ST_conf_factor}) $t=0$, and ending at time $t=42\sqrt{GM}$.  At late times, our numerical results for $\gamma_{RR}$ agree very well with the analytical results of Section \ref{1pluslogDeq5subsect}.  The latter are included as the solid line, which overlaps with the line representing the numerical results at $t = 42 \sqrt{GM}$.  }
\label{Fig:psi_1+log}
\end{figure}

\begin{figure}
\begin{center}
\includegraphics[width=3.25in]{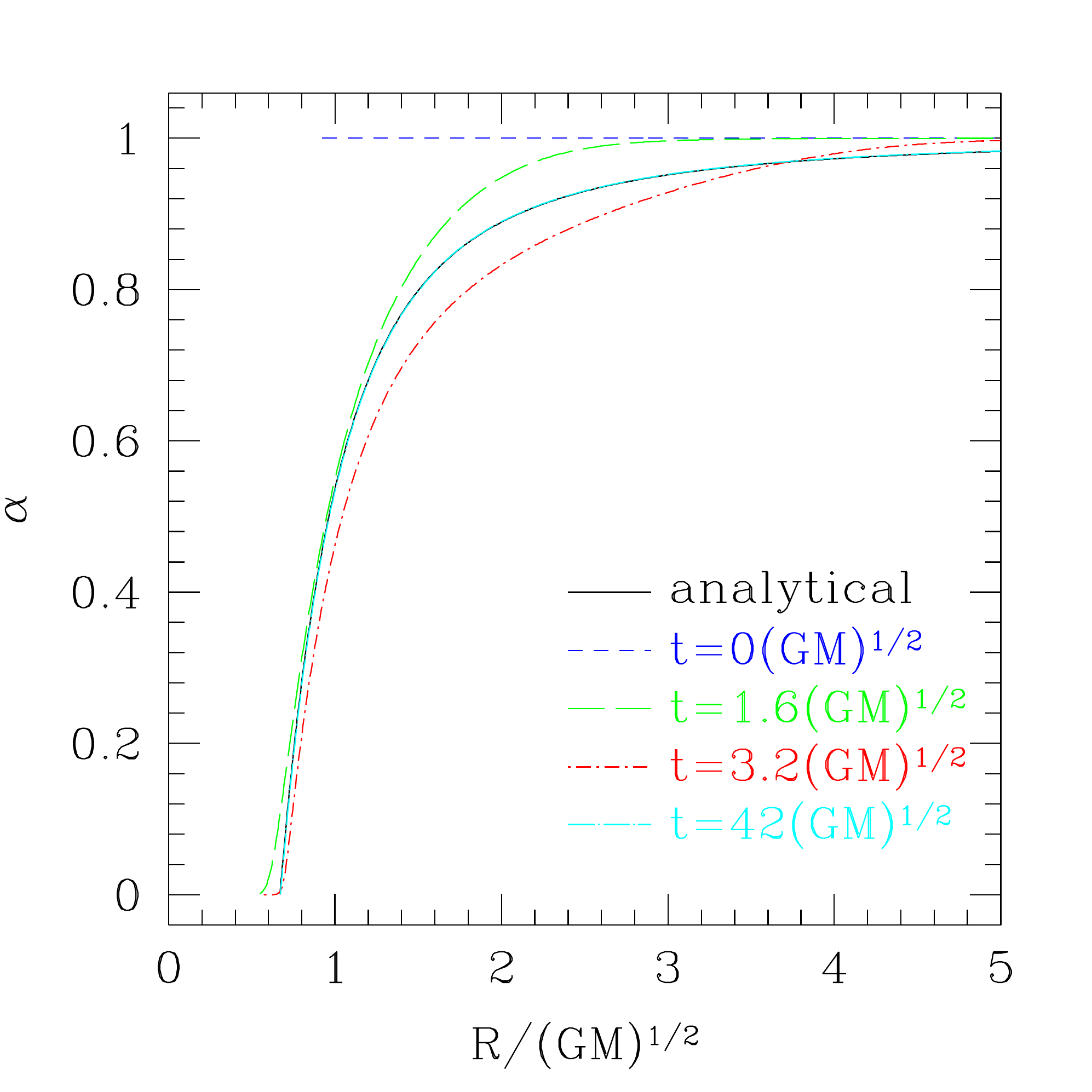}
\end{center}
\caption{Same as Fig.~\ref{Fig:psi_1+log}, but for the lapse $\alpha$.}
\label{Fig:alpha_1+log}
\end{figure}

\begin{figure} 
\begin{center}
\includegraphics[width=3.25in]{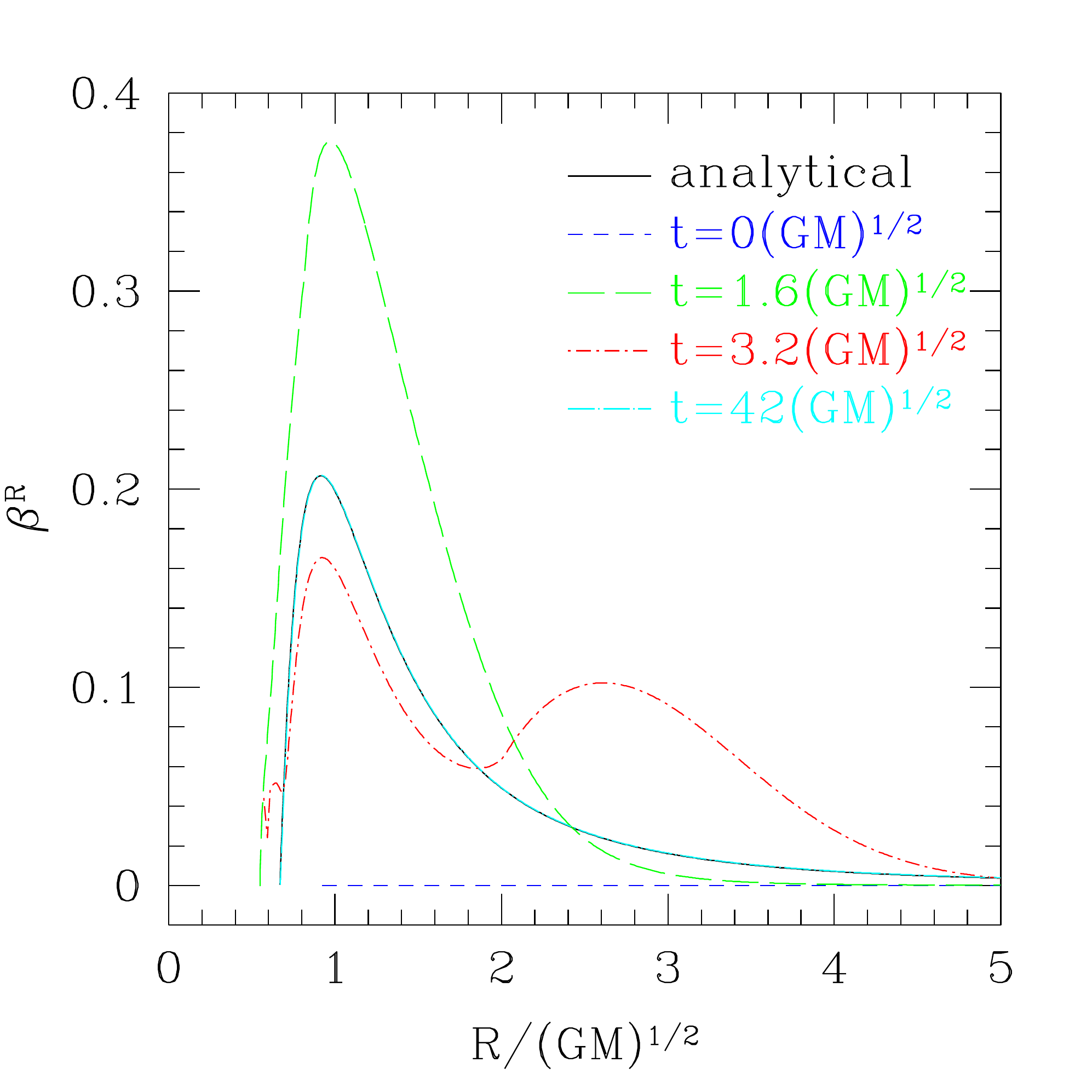}
\end{center}
\caption{Same as Fig.~\ref{Fig:psi_1+log}, but for the shift $\beta^R$.}
\label{Fig:shift_1+log}
\end{figure}

In the following figures we graph several quantities in terms of the areal radius $R$, which simplifies comparison with the analytical results for the late-time asymptotic solutions.  In terms of the variables evolved in our code, we compute $R$ from $R=\sqrt{\bar \gamma_{\theta\theta}/\chi}$, where $\bar \gamma_{\theta\theta}$ is the $\theta\theta$ component of the conformally related metric and $\chi=\psi^{-2}$. 

The initial data for our dynamical simulations are given in terms of an isotropic spatial metric.  However, this spatial metric is evolved in time, and does not remain isotropic (see Figs.~35 and 36 in \cite{Bro09}), so that we cannot compare the conformal factor in the code directly with the conformal factor that we would obtain by transforming the analytical results of the previous sections into isotropic coordinates.  Instead, we compare analytical and numerical values for the metric component $\gamma_{RR}$ in Fig.~\ref{Fig:psi_1+log}.  We compute numerical values for $\gamma_{RR}$ from 
\begin{equation}
\gamma_{RR}=\frac{\bar \gamma_{rr}}{\chi\left(dR/dr \right)^{2}},
\end{equation}
where $\bar \gamma_{rr}$ is the 
$rr$ component of the conformally related metric and the derivative $dR/dr$ is calculated numerically from the areal radius $R$ given above. 

We show the evolution of the numerical $\gamma_{RR}$ at several different times, starting with the initial data (\ref{ST_conf_factor}) at $t=0$  and ending at a time $t=42\sqrt{GM}$ before the results shown in the figure are affected by the presence of the outer boundary.   (In order to avoid double-valued functions, we show the numerical initial data only for isotropic radii outside the event horizon.)  Also included are the analytical results obtained in Section \ref{1pluslogDeq5subsect}.  
We see that $\gamma_{RR}$ evolves for a certain period of time, and then settles down to the expected time-independent solution.

In Figs.~\ref{Fig:alpha_1+log} and \ref{Fig:shift_1+log} we show similar results for the lapse and the shift.  The numerical values of the areal shift $\beta^{R}$ are calculated using $\beta^{R}=(dR/dr)\beta^{r}$.  Both the lapse and shift start with their initial values (\ref{alpha_beta_init}), go through a dynamical phase, and then settle down to their asymptotic values.  At these late times we find excellent agreement with the stationary solutions that we computed analytically in Section \ref{1pluslogDeq5subsect}.  

\subsubsection{Maximal slicing}

\begin{figure} 
\begin{center}
\includegraphics[width=3.25in]{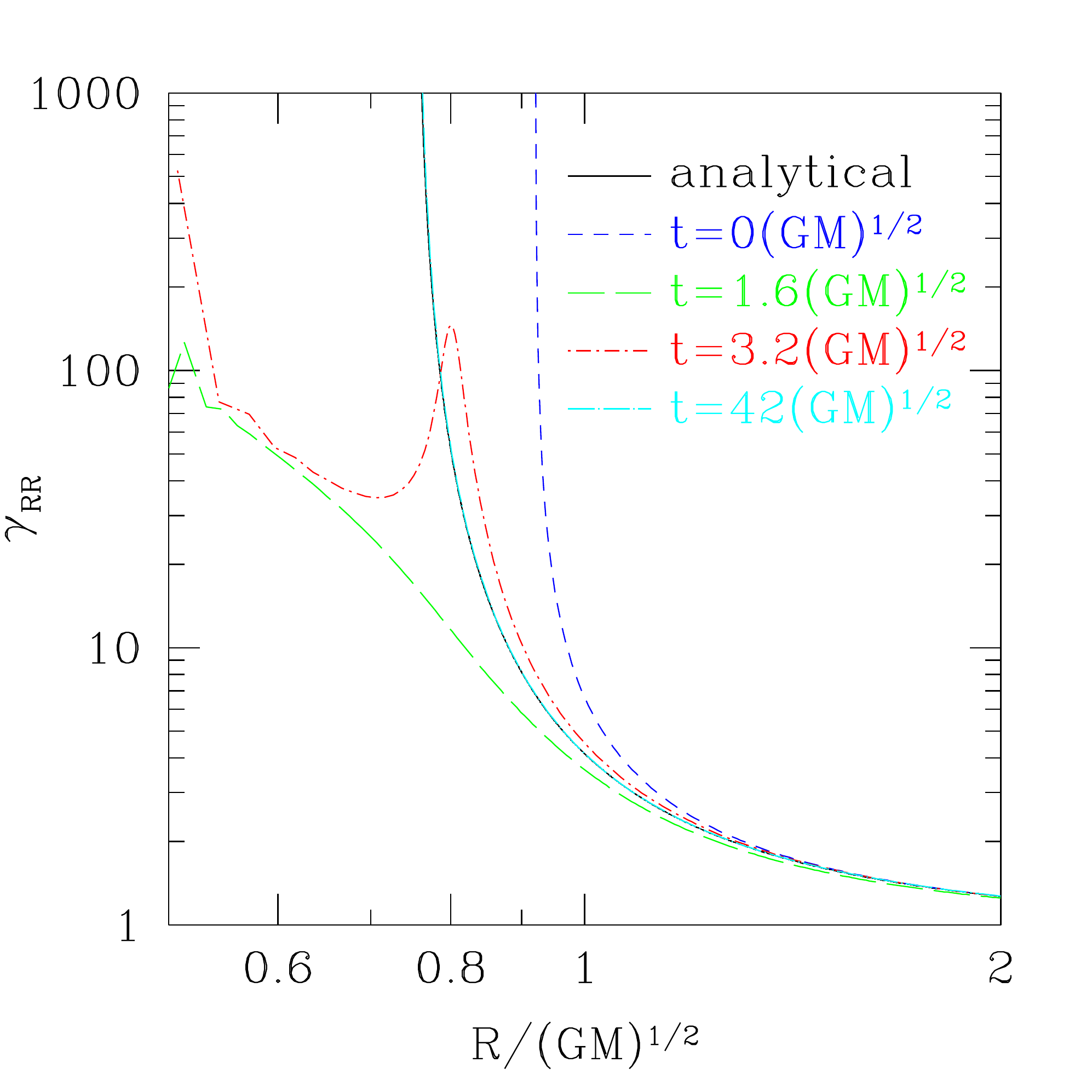}
\end{center}
\caption{The metric component $\gamma_{RR}$ as a function of the areal radius $R$ for a dynamical evolution of the $D=5$ Schwarzschild-Tangherlini spacetime with the non-advective 1+log slicing condition (\ref{1+log_na}) for $n=2$.  We show $\gamma_{RR}$ at a number of different times, starting with the initial data (\ref{ST_conf_factor}) $t=0$, and ending at time $t=42\sqrt{GM}$.  At late times, $\gamma_{RR}$ approaches the trumpet member of the family of maximal slices discussed in Section \ref{maxDeq5subsect}.  The latter is included as the solid line, which overlaps the line representing the numerical results at $t = 42 \sqrt{GM}$.}
\label{Fig:psi_1+log_na}
\end{figure}

We also performed numerical simulations for the ``non-advective" 1+log slicing condition, for which the advective term in (\ref{1+log}) is dropped,
\begin{equation} \label{1+log_na}
\partial_t \alpha = - n \alpha K.
\end{equation}
Time-independent solutions must satisfy the maximal slicing condition, $K=0$.  If dynamical simulations with this slicing condition settle down to a time-independent solution, then the late-time asymptotic solution must be given by a member of the family of maximal slices that we discussed in Section \ref{MaxSlicSection}.  As it turns out, dynamical simulations settle down to the ``special" member with $C = \tilde C$, which displays a trumpet geometry (see also the discussion in \cite{HanHOBGS06}).   We demonstrate this behavior in Fig.~\ref{Fig:psi_1+log_na}, where we test the time evolution of the conformal factor $\psi$ by plotting $\gamma_{RR}$ as a function of areal radius $R$.  As before, the evolution starts with the initial data (\ref{ST_conf_factor}), but now the evolution settles down to the trumpet member of the family of maximal slices discussed in Section \ref{maxDeq5subsect} (see also \cite{NakAYS09}).

%=====================================================================

\section{Summary}
\label{Summary}

We study maximal and stationary 1+log slices of the Schwarzschild-Tangherlini spacetime for $D \geq 4$ spacetime dimensions.  We use a height function to introduce a coordinate transformation away from slices of constant Schwarzschild-Tangherlini time, and find families of both maximal slices and stationary 1+log slices.  

For the maximal slices, which were previously considered by \cite{NakAYS09}, we identify one ``special" member that displays a trumpet geometry.  For the stationary 1+log slices we impose a regularity condition following the methods of \cite{HanHOBO08}.  This regularity condition singles out one particular member, which also displays a trumpet geometry.  We allow for a free parameter $n$ in our 1+log slicing condition, where the limit $n \rightarrow \infty$ corresponds to maximal slicing.  We demonstrate that our results for 1+log slicing reduce to the maximal slicing results of \cite{NakAYS09} in this limit, and we also show that we recover the results of \cite{HanHOBO08} for $D=4$.  

Finally, we perform numerical simulations for spherically symmetric black holes in $D=5$ spacetime dimensions.  We start with data on a slice of constant Schwarzschild-Tangherlini time, and evolve these data with moving-puncture gauge conditions.  Our results demonstrate that, as for $D=4$ spacetime dimensions, the dynamical simulations settle down to the trumpet slices, which can be regarded as ``attractors" for moving-puncture simulations.

%=====================================================================

\begin{acknowledgments}
It is a pleasure to thank Alexa N. Staley for her assistance with the numerical part of this project. TWB would like to thank S.~Husa for helpful conversations.  JPW gratefully acknowledges support through a Gibbons Undergraduate Fellowship and from the Maine Space Grant Consortium. This work was supported in part by NSF grant PHY-0756514 to Bowdoin College and by NSF grant PHY-0758116 to North Carolina State University.
\end{acknowledgments}

% \bibliography{ref.bib}

\begin{thebibliography}{34}
\expandafter\ifx\csname natexlab\endcsname\relax\def\natexlab#1{#1}\fi
\expandafter\ifx\csname bibnamefont\endcsname\relax
  \def\bibnamefont#1{#1}\fi
\expandafter\ifx\csname bibfnamefont\endcsname\relax
  \def\bibfnamefont#1{#1}\fi
\expandafter\ifx\csname citenamefont\endcsname\relax
  \def\citenamefont#1{#1}\fi
\expandafter\ifx\csname url\endcsname\relax
  \def\url#1{\texttt{#1}}\fi
\expandafter\ifx\csname urlprefix\endcsname\relax\def\urlprefix{URL }\fi
\providecommand{\bibinfo}[2]{#2}
\providecommand{\eprint}[2][]{\url{#2}}

\bibitem[{\citenamefont{{Pretorius}}(2005)}]{Pre05b}
\bibinfo{author}{\bibfnamefont{F.}~\bibnamefont{{Pretorius}}},
  \bibinfo{journal}{\prl} \textbf{\bibinfo{volume}{95}},
  \bibinfo{pages}{121101/1} (\bibinfo{year}{2005}).

\bibitem[{\citenamefont{{Campanelli} et~al.}(2006)\citenamefont{{Campanelli},
  {Lousto}, {Marronetti}, and {Zlochower}}}]{CamLMZ06}
\bibinfo{author}{\bibfnamefont{M.}~\bibnamefont{{Campanelli}}},
  \bibinfo{author}{\bibfnamefont{C.~O.} \bibnamefont{{Lousto}}},
  \bibinfo{author}{\bibfnamefont{P.}~\bibnamefont{{Marronetti}}},
  \bibnamefont{and}
  \bibinfo{author}{\bibfnamefont{Y.}~\bibnamefont{{Zlochower}}},
  \bibinfo{journal}{\prl} \textbf{\bibinfo{volume}{96}},
  \bibinfo{pages}{111101/1} (\bibinfo{year}{2006}).

\bibitem[{\citenamefont{{Baker} et~al.}({2006})\citenamefont{{Baker},
  {Centrella}, {Choi}, {Koppitz}, and {van Meter}}}]{BakCCKM06a}
\bibinfo{author}{\bibfnamefont{J.~G.} \bibnamefont{{Baker}}},
  \bibinfo{author}{\bibfnamefont{J.}~\bibnamefont{{Centrella}}},
  \bibinfo{author}{\bibfnamefont{D.-I.} \bibnamefont{{Choi}}},
  \bibinfo{author}{\bibfnamefont{M.}~\bibnamefont{{Koppitz}}},
  \bibnamefont{and} \bibinfo{author}{\bibfnamefont{J.}~\bibnamefont{{van
  Meter}}}, \bibinfo{journal}{\prl} \textbf{\bibinfo{volume}{96}},
  \bibinfo{pages}{111102/1} (\bibinfo{year}{{2006}}).

\bibitem[{\citenamefont{{Campanelli} et~al.}(2007)\citenamefont{{Campanelli},
  {Lousto}, {Zlochower}, and Merritt}}]{CamLZM07b}
\bibinfo{author}{\bibfnamefont{M.}~\bibnamefont{{Campanelli}}},
  \bibinfo{author}{\bibfnamefont{C.~O.} \bibnamefont{{Lousto}}},
  \bibinfo{author}{\bibfnamefont{Y.}~\bibnamefont{{Zlochower}}},
  \bibnamefont{and} \bibinfo{author}{\bibfnamefont{D.}~\bibnamefont{Merritt}},
  \bibinfo{journal}{\prl} \textbf{\bibinfo{volume}{98}},
  \bibinfo{pages}{231102/1} (\bibinfo{year}{2007}).

\bibitem[{\citenamefont{{Gonz\'alez} et~al.}(2007)\citenamefont{{Gonz\'alez},
  {Hannam}, {Sperhake}, {Brugmann}, and {Husa}}}]{GonHSBH07}
\bibinfo{author}{\bibfnamefont{J.~A.} \bibnamefont{{Gonz\'alez}}},
  \bibinfo{author}{\bibfnamefont{M.~D.} \bibnamefont{{Hannam}}},
  \bibinfo{author}{\bibfnamefont{U.}~\bibnamefont{{Sperhake}}},
  \bibinfo{author}{\bibfnamefont{B.}~\bibnamefont{{Brugmann}}},
  \bibnamefont{and} \bibinfo{author}{\bibfnamefont{S.}~\bibnamefont{{Husa}}},
  \bibinfo{journal}{\prl} \textbf{\bibinfo{volume}{98}},
  \bibinfo{pages}{231101/1} (\bibinfo{year}{2007}).

\bibitem[{\citenamefont{{Shibata} and {Nakamura}}(1995)}]{ShiN95}
\bibinfo{author}{\bibfnamefont{M.}~\bibnamefont{{Shibata}}} \bibnamefont{and}
  \bibinfo{author}{\bibfnamefont{T.}~\bibnamefont{{Nakamura}}},
  \bibinfo{journal}{\prd} \textbf{\bibinfo{volume}{52}}, \bibinfo{pages}{5428}
  (\bibinfo{year}{1995}).

\bibitem[{\citenamefont{{Baumgarte} and {Shapiro}}(1998)}]{BauS99}
\bibinfo{author}{\bibfnamefont{T.~W.} \bibnamefont{{Baumgarte}}}
  \bibnamefont{and} \bibinfo{author}{\bibfnamefont{S.~L.}
  \bibnamefont{{Shapiro}}}, \bibinfo{journal}{\prd}
  \textbf{\bibinfo{volume}{59}}, \bibinfo{pages}{024007/1}
  (\bibinfo{year}{1998}).

\bibitem[{\citenamefont{{Baumgarte} and {Shapiro}}({2010})}]{BauS10}
\bibinfo{author}{\bibfnamefont{T.~W.} \bibnamefont{{Baumgarte}}}
  \bibnamefont{and} \bibinfo{author}{\bibfnamefont{S.~L.}
  \bibnamefont{{Shapiro}}}, \emph{\bibinfo{title}{Numerical Relativity: Solving
  Einstein's Equations on the Computer}} (\bibinfo{publisher}{Cambridge
  University Press, Cambridge}, \bibinfo{year}{{2010}}).

\bibitem[{\citenamefont{{Bona} et~al.}(1995)\citenamefont{{Bona}, {Mass{\'o}},
  {Seidel}, and {Stela}}}]{BonMSS95}
\bibinfo{author}{\bibfnamefont{C.}~\bibnamefont{{Bona}}},
  \bibinfo{author}{\bibfnamefont{J.}~\bibnamefont{{Mass{\'o}}}},
  \bibinfo{author}{\bibfnamefont{E.}~\bibnamefont{{Seidel}}}, \bibnamefont{and}
  \bibinfo{author}{\bibfnamefont{J.}~\bibnamefont{{Stela}}},
  \bibinfo{journal}{\prl} \textbf{\bibinfo{volume}{75}}, \bibinfo{pages}{600}
  (\bibinfo{year}{1995}).

\bibitem[{\citenamefont{{Alcubierre} et~al.}(2003)\citenamefont{{Alcubierre},
  {Br{\"u}gmann}, {Diener}, {Koppitz}, {Pollney}, {Seidel}, and
  {Takahashi}}}]{AlcBDKPST03}
\bibinfo{author}{\bibfnamefont{M.}~\bibnamefont{{Alcubierre}}},
  \bibinfo{author}{\bibfnamefont{B.}~\bibnamefont{{Br{\"u}gmann}}},
  \bibinfo{author}{\bibfnamefont{P.}~\bibnamefont{{Diener}}},
  \bibinfo{author}{\bibfnamefont{M.}~\bibnamefont{{Koppitz}}},
  \bibinfo{author}{\bibfnamefont{D.}~\bibnamefont{{Pollney}}},
  \bibinfo{author}{\bibfnamefont{E.}~\bibnamefont{{Seidel}}}, \bibnamefont{and}
  \bibinfo{author}{\bibfnamefont{R.}~\bibnamefont{{Takahashi}}},
  \bibinfo{journal}{\prd} \textbf{\bibinfo{volume}{67}},
  \bibinfo{pages}{084023/1} (\bibinfo{year}{2003}).

\bibitem[{\citenamefont{{Hannam}
  et~al.}(2007{\natexlab{a}})\citenamefont{{Hannam}, {Husa}, {Pollney},
  {Bruegmann}, and {O'Murchadha}}}]{HanHPBO06}
\bibinfo{author}{\bibfnamefont{M.}~\bibnamefont{{Hannam}}},
  \bibinfo{author}{\bibfnamefont{S.}~\bibnamefont{{Husa}}},
  \bibinfo{author}{\bibfnamefont{D.}~\bibnamefont{{Pollney}}},
  \bibinfo{author}{\bibfnamefont{B.}~\bibnamefont{{Bruegmann}}},
  \bibnamefont{and}
  \bibinfo{author}{\bibfnamefont{N.}~\bibnamefont{{O'Murchadha}}},
  \bibinfo{journal}{\prl} \textbf{\bibinfo{volume}{99}},
  \bibinfo{pages}{241102/1} (\bibinfo{year}{2007}{\natexlab{a}}).

\bibitem[{\citenamefont{Brown}(2008)}]{Bro08}
\bibinfo{author}{\bibfnamefont{J.~D.} \bibnamefont{Brown}},
  \bibinfo{journal}{\prd} \textbf{\bibinfo{volume}{77}},
  \bibinfo{pages}{044018/1} (\bibinfo{year}{2008}).

\bibitem[{\citenamefont{{Hannam} et~al.}(2008)\citenamefont{{Hannam}, {Husa},
  {Ohme}, {Br{\"u}gmann}, and {Murchadha}}}]{HanHOBO08}
\bibinfo{author}{\bibfnamefont{M.}~\bibnamefont{{Hannam}}},
  \bibinfo{author}{\bibfnamefont{S.}~\bibnamefont{{Husa}}},
  \bibinfo{author}{\bibfnamefont{F.}~\bibnamefont{{Ohme}}},
  \bibinfo{author}{\bibfnamefont{B.}~\bibnamefont{{Br{\"u}gmann}}},
  \bibnamefont{and} \bibinfo{author}{\bibfnamefont{N.~{\'O}.}
  \bibnamefont{{Murchadha}}}, \bibinfo{journal}{\prd}
  \textbf{\bibinfo{volume}{78}}, \bibinfo{pages}{064020/1}
  (\bibinfo{year}{2008}).

\bibitem[{\citenamefont{{Choptuik} et~al.}(2003)\citenamefont{{Choptuik},
  {Lehner}, {Olabarrieta}, {Petryk}, {Pretorius}, and {Villegas}}}]{ChoLOPPV03}
\bibinfo{author}{\bibfnamefont{M.}~\bibnamefont{{Choptuik}}},
  \bibinfo{author}{\bibfnamefont{L.}~\bibnamefont{{Lehner}}},
  \bibinfo{author}{\bibfnamefont{I.}~\bibnamefont{{Olabarrieta}}},
  \bibinfo{author}{\bibfnamefont{R.}~\bibnamefont{{Petryk}}},
  \bibinfo{author}{\bibfnamefont{F.}~\bibnamefont{{Pretorius}}},
  \bibnamefont{and}
  \bibinfo{author}{\bibfnamefont{H.}~\bibnamefont{{Villegas}}},
  \bibinfo{journal}{\prd} \textbf{\bibinfo{volume}{68}},
  \bibinfo{pages}{044001/1} (\bibinfo{year}{2003}).

\bibitem[{\citenamefont{{Garfinkle} et~al.}(2005)\citenamefont{{Garfinkle},
  {Lehner}, and {Pretorius}}}]{GarLP05}
\bibinfo{author}{\bibfnamefont{D.}~\bibnamefont{{Garfinkle}}},
  \bibinfo{author}{\bibfnamefont{L.}~\bibnamefont{{Lehner}}}, \bibnamefont{and}
  \bibinfo{author}{\bibfnamefont{F.}~\bibnamefont{{Pretorius}}},
  \bibinfo{journal}{\prd} \textbf{\bibinfo{volume}{71}},
  \bibinfo{pages}{064009/1} (\bibinfo{year}{2005}).

\bibitem[{\citenamefont{{Lehner} and {Pretorius}}(2010)}]{LehP10}
\bibinfo{author}{\bibfnamefont{L.}~\bibnamefont{{Lehner}}} \bibnamefont{and}
  \bibinfo{author}{\bibfnamefont{F.}~\bibnamefont{{Pretorius}}}
  (\bibinfo{year}{2010}), \eprint{arXiv:1006.5960}.

\bibitem[{\citenamefont{{Yoshino} and {Shibata}}(2009)}]{YosS09}
\bibinfo{author}{\bibfnamefont{H.}~\bibnamefont{{Yoshino}}} \bibnamefont{and}
  \bibinfo{author}{\bibfnamefont{M.}~\bibnamefont{{Shibata}}},
  \bibinfo{journal}{\prd} \textbf{\bibinfo{volume}{80}},
  \bibinfo{pages}{084025/1} (\bibinfo{year}{2009}).

\bibitem[{\citenamefont{{Nakao} et~al.}(2009)\citenamefont{{Nakao}, {Abe},
  {Yoshino}, and {Shibata}}}]{NakAYS09}
\bibinfo{author}{\bibfnamefont{K.-i.} \bibnamefont{{Nakao}}},
  \bibinfo{author}{\bibfnamefont{H.}~\bibnamefont{{Abe}}},
  \bibinfo{author}{\bibfnamefont{H.}~\bibnamefont{{Yoshino}}},
  \bibnamefont{and}
  \bibinfo{author}{\bibfnamefont{M.}~\bibnamefont{{Shibata}}},
  \bibinfo{journal}{\prd} \textbf{\bibinfo{volume}{80}},
  \bibinfo{pages}{084028/1} (\bibinfo{year}{2009}).

\bibitem[{\citenamefont{{Shibata} and {Yoshino}}(2010{\natexlab{a}})}]{ShiY10a}
\bibinfo{author}{\bibfnamefont{M.}~\bibnamefont{{Shibata}}} \bibnamefont{and}
  \bibinfo{author}{\bibfnamefont{H.}~\bibnamefont{{Yoshino}}},
  \bibinfo{journal}{\prd} \textbf{\bibinfo{volume}{81}},
  \bibinfo{pages}{021501(R)/1} (\bibinfo{year}{2010}{\natexlab{a}}).

\bibitem[{\citenamefont{{Shibata} and {Yoshino}}(2010{\natexlab{b}})}]{ShiY10b}
\bibinfo{author}{\bibfnamefont{M.}~\bibnamefont{{Shibata}}} \bibnamefont{and}
  \bibinfo{author}{\bibfnamefont{H.}~\bibnamefont{{Yoshino}}},
  \bibinfo{journal}{\prd} \textbf{\bibinfo{volume}{81}},
  \bibinfo{pages}{104035/1} (\bibinfo{year}{2010}{\natexlab{b}}).

\bibitem[{\citenamefont{{Zilh\~{a}o} et~al.}(2010)\citenamefont{{Zilh\~{a}o},
  {Witek}, {Sperhake}, {Cardoso}, {Gualtieri}, {Herdeiro}, and
  {Nerozzi}}}]{ZilWSCGHN10}
\bibinfo{author}{\bibfnamefont{M.}~\bibnamefont{{Zilh\~{a}o}}},
  \bibinfo{author}{\bibfnamefont{H.}~\bibnamefont{{Witek}}},
  \bibinfo{author}{\bibfnamefont{U.}~\bibnamefont{{Sperhake}}},
  \bibinfo{author}{\bibfnamefont{V.}~\bibnamefont{{Cardoso}}},
  \bibinfo{author}{\bibfnamefont{L.}~\bibnamefont{{Gualtieri}}},
  \bibinfo{author}{\bibfnamefont{C.}~\bibnamefont{{Herdeiro}}},
  \bibnamefont{and}
  \bibinfo{author}{\bibfnamefont{A.}~\bibnamefont{{Nerozzi}}},
  \bibinfo{journal}{\prd} \textbf{\bibinfo{volume}{81}},
  \bibinfo{pages}{084052/1} (\bibinfo{year}{2010}).

\bibitem[{\citenamefont{{Witek} et~al.}(2010)\citenamefont{{Witek},
  {Zilh\~{a}o}, {Gualtieri}, {Cardoso}, {Herdeiro}, {Nerozzi}, and
  {Sperhake}}}]{WitZGCHNS10}
\bibinfo{author}{\bibfnamefont{H.}~\bibnamefont{{Witek}}},
  \bibinfo{author}{\bibfnamefont{M.}~\bibnamefont{{Zilh\~{a}o}}},
  \bibinfo{author}{\bibfnamefont{L.}~\bibnamefont{{Gualtieri}}},
  \bibinfo{author}{\bibfnamefont{V.}~\bibnamefont{{Cardoso}}},
  \bibinfo{author}{\bibfnamefont{C.}~\bibnamefont{{Herdeiro}}},
  \bibinfo{author}{\bibfnamefont{A.}~\bibnamefont{{Nerozzi}}},
  \bibnamefont{and}
  \bibinfo{author}{\bibfnamefont{U.}~\bibnamefont{{Sperhake}}}
  (\bibinfo{year}{2010}), \eprint{arXiv:1006.3081}.

\bibitem[{\citenamefont{{Emparan} and {Reall}}(2008)}]{EmpR08}
\bibinfo{author}{\bibfnamefont{R.}~\bibnamefont{{Emparan}}} \bibnamefont{and}
  \bibinfo{author}{\bibfnamefont{H.~S.} \bibnamefont{{Reall}}},
  \bibinfo{journal}{Living Reviews in Relativity} \textbf{\bibinfo{volume}{11}}
  (\bibinfo{year}{2008}),
  \urlprefix\url{http://www.livingreviews.org/lrr-2008-6}.

\bibitem[{\citenamefont{{Kanti}}(2009)}]{Kan09}
\bibinfo{author}{\bibfnamefont{P.}~\bibnamefont{{Kanti}}},
  \bibinfo{journal}{Lect. Notes Phys.} \textbf{\bibinfo{volume}{769}},
  \bibinfo{pages}{387} (\bibinfo{year}{2009}).

\bibitem[{\citenamefont{{Tangherlini}}(1963)}]{Tan63}
\bibinfo{author}{\bibfnamefont{F.~R.} \bibnamefont{{Tangherlini}}},
  \bibinfo{journal}{Il Nuovo Cimento} \textbf{\bibinfo{volume}{27}},
  \bibinfo{pages}{636} (\bibinfo{year}{1963}).

\bibitem[{\citenamefont{{Kol} et~al.}(2004)\citenamefont{{Kol}, {Sorkin}, and
  {Piran}}}]{KolSP04}
\bibinfo{author}{\bibfnamefont{B.}~\bibnamefont{{Kol}}},
  \bibinfo{author}{\bibfnamefont{E.}~\bibnamefont{{Sorkin}}}, \bibnamefont{and}
  \bibinfo{author}{\bibfnamefont{T.}~\bibnamefont{{Piran}}},
  \bibinfo{journal}{\prd} \textbf{\bibinfo{volume}{69}},
  \bibinfo{pages}{064031/1} (\bibinfo{year}{2004}).

\bibitem[{\citenamefont{{Reinhart}}(1973)}]{Rei73}
\bibinfo{author}{\bibfnamefont{B.}~\bibnamefont{{Reinhart}}},
  \bibinfo{journal}{J. Math. Phys.} \textbf{\bibinfo{volume}{14}},
  \bibinfo{pages}{719} (\bibinfo{year}{1973}).

\bibitem[{\citenamefont{{Estabrook} et~al.}(1973)\citenamefont{{Estabrook},
  {Wahlquist}, {Christensen}, {DeWitt}, {Smarr}, and {Tsiang}}}]{EstWCDST}
\bibinfo{author}{\bibfnamefont{F.}~\bibnamefont{{Estabrook}}},
  \bibinfo{author}{\bibfnamefont{H.}~\bibnamefont{{Wahlquist}}},
  \bibinfo{author}{\bibfnamefont{S.}~\bibnamefont{{Christensen}}},
  \bibinfo{author}{\bibfnamefont{B.}~\bibnamefont{{DeWitt}}},
  \bibinfo{author}{\bibfnamefont{L.}~\bibnamefont{{Smarr}}}, \bibnamefont{and}
  \bibinfo{author}{\bibfnamefont{E.}~\bibnamefont{{Tsiang}}},
  \bibinfo{journal}{\prd} \textbf{\bibinfo{volume}{7}}, \bibinfo{pages}{2814}
  (\bibinfo{year}{1973}).

\bibitem[{\citenamefont{{Beig} and {{\'O} Murchadha}}(1998)}]{BeiO98}
\bibinfo{author}{\bibfnamefont{R.}~\bibnamefont{{Beig}}} \bibnamefont{and}
  \bibinfo{author}{\bibfnamefont{N.}~\bibnamefont{{{\'O} Murchadha}}},
  \bibinfo{journal}{\prd} \textbf{\bibinfo{volume}{57}}, \bibinfo{pages}{4728}
  (\bibinfo{year}{1998}).

\bibitem[{\citenamefont{{Baumgarte} and {Naculich}}(2007)}]{BauN07}
\bibinfo{author}{\bibfnamefont{T.~W.} \bibnamefont{{Baumgarte}}}
  \bibnamefont{and} \bibinfo{author}{\bibfnamefont{S.~G.}
  \bibnamefont{{Naculich}}}, \bibinfo{journal}{\prd}
  \textbf{\bibinfo{volume}{75}}, \bibinfo{pages}{067502/1}
  (\bibinfo{year}{2007}).

\bibitem[{\citenamefont{Brown}(2009)}]{Bro09}
\bibinfo{author}{\bibfnamefont{J.~D.} \bibnamefont{Brown}},
  \bibinfo{journal}{\prd} \textbf{\bibinfo{volume}{80}},
  \bibinfo{pages}{084042/1} (\bibinfo{year}{2009}).

\bibitem[{\citenamefont{Alcubierre et~al.}(2001)\citenamefont{Alcubierre,
  Brandt, Br\"{u}gmann, Holz, Seidel, Takahashi, and Thornburg}}]{AlcBBHSTT01}
\bibinfo{author}{\bibfnamefont{M.}~\bibnamefont{Alcubierre}},
  \bibinfo{author}{\bibfnamefont{S.}~\bibnamefont{Brandt}},
  \bibinfo{author}{\bibfnamefont{B.}~\bibnamefont{Br\"{u}gmann}},
  \bibinfo{author}{\bibfnamefont{D.}~\bibnamefont{Holz}},
  \bibinfo{author}{\bibfnamefont{E.}~\bibnamefont{Seidel}},
  \bibinfo{author}{\bibfnamefont{R.}~\bibnamefont{Takahashi}},
  \bibnamefont{and}
  \bibinfo{author}{\bibfnamefont{J.}~\bibnamefont{Thornburg}},
  \bibinfo{journal}{Int.~J.~Mod.~Phys.~D} \textbf{\bibinfo{volume}{10}},
  \bibinfo{pages}{273} (\bibinfo{year}{2001}).

\bibitem[{\citenamefont{{Alcubierre} et~al.}(2005)\citenamefont{{Alcubierre},
  {Br{\"u}gmann}, {Diener}, {Guzm{\'a}n}, {Hawke}, {Hawley}, {Herrmann},
  {Koppitz}, {Pollney}, {Seidel} et~al.}}]{AlcBDGHHHKPST05}
\bibinfo{author}{\bibfnamefont{M.}~\bibnamefont{{Alcubierre}}},
  \bibinfo{author}{\bibfnamefont{B.}~\bibnamefont{{Br{\"u}gmann}}},
  \bibinfo{author}{\bibfnamefont{P.}~\bibnamefont{{Diener}}},
  \bibinfo{author}{\bibfnamefont{F.~S.} \bibnamefont{{Guzm{\'a}n}}},
  \bibinfo{author}{\bibfnamefont{I.}~\bibnamefont{{Hawke}}},
  \bibinfo{author}{\bibfnamefont{S.}~\bibnamefont{{Hawley}}},
  \bibinfo{author}{\bibfnamefont{F.}~\bibnamefont{{Herrmann}}},
  \bibinfo{author}{\bibfnamefont{M.}~\bibnamefont{{Koppitz}}},
  \bibinfo{author}{\bibfnamefont{D.}~\bibnamefont{{Pollney}}},
  \bibinfo{author}{\bibfnamefont{E.}~\bibnamefont{{Seidel}}},
  \bibnamefont{et~al.}, \bibinfo{journal}{\prd} \textbf{\bibinfo{volume}{72}},
  \bibinfo{pages}{044004/1} (\bibinfo{year}{2005}).

\bibitem[{\citenamefont{{Hannam}
  et~al.}(2007{\natexlab{b}})\citenamefont{{Hannam}, {Husa}, {Murchadha},
  {Br{\"u}gmann}, {Gonz{\'a}lez}, and {Sperhake}}}]{HanHOBGS06}
\bibinfo{author}{\bibfnamefont{M.}~\bibnamefont{{Hannam}}},
  \bibinfo{author}{\bibfnamefont{S.}~\bibnamefont{{Husa}}},
  \bibinfo{author}{\bibfnamefont{N.~{\'O}.} \bibnamefont{{Murchadha}}},
  \bibinfo{author}{\bibfnamefont{B.}~\bibnamefont{{Br{\"u}gmann}}},
  \bibinfo{author}{\bibfnamefont{J.~A.} \bibnamefont{{Gonz{\'a}lez}}},
  \bibnamefont{and}
  \bibinfo{author}{\bibfnamefont{U.}~\bibnamefont{{Sperhake}}},
  \bibinfo{journal}{{J.~Phys.~Conf.~Series}} \textbf{\bibinfo{volume}{66}},
  \bibinfo{pages}{012047/1} (\bibinfo{year}{2007}{\natexlab{b}}).

\end{thebibliography}

\end{document}